\documentclass[twocolumn,showpacs,superscriptaddress,amsmath,amssymb,prb]{revtex4-1}

\usepackage[dvips]{graphicx}% Include figure files
\usepackage{dcolumn}% Align table columns on decimal point
\usepackage{bm}% bold math

\begin{document}

%\title{Magnetic field and temperature effects on CuO chain superconductivity in YBa$_{2}$Cu$_{4}$O$_{8}$}
%\title{Magnetic field-dependence of the basal-plane superconducting anisotropy in double-chained YBa$_{2}$Cu$_{4}$O$_{8}$ from small-angle neutron scattering measurements of the vortex lattice}

\title{Magnetic field-dependence of the basal-plane superconducting anisotropy in YBa$_{2}$Cu$_{4}$O$_{8}$ from small-angle neutron scattering measurements of the vortex lattice}

\author{Jonathan\,S.\,White,$^{1-3}$ Charlotte\,J.\,Bowell,$^{2}$ Alistair\,S.\,Cameron,$^{2}$ Richard\,W.\,Heslop,$^{2}$ Jo\"{e}l\,Mesot,$^{1,3,4}$ Jorge\,L.\,Gavilano,$^{1}$ Simon\,Str\"{a}ssle,$^{5}$ Lars\,M\"{a}chler,$^{1}$ Rustem\,Khasanov,$^{6}$ Charles\,D.\,Dewhurst,$^{7}$ Janusz\,Karpinski,$^{4}$ and Edward\,M.\,Forgan$^{2}$}

\affiliation{Laboratory for Neutron Scattering, Paul Scherrer Institut, CH-5232 Villigen, Switzerland\\
$^{2}$School of Physics and Astronomy, University of Birmingham, Edgbaston, Birmingham, B15 2TT, UK\\
$^{3}$Institute of Condensed Matter Physics, Ecole Polytechnique F\'{e}d\'{e}rale de Lausanne, CH-1015 Lausanne, Switzerland\\
$^{4}$Laboratory for Solid State Physics, ETH Z\"{u}rich, CH-8093 Z\"{u}rich, Switzerland\\
$^{5}$Physik-Institut der Universit\"{a}t Z\"{u}rich, Winterthurerstrasse 190, CH-8057 Z\"{u}rich, Switzerland\\
$^{6}$Laboratory for Muon Spin Spectroscopy, Paul Scherrer Institute, CH-5232 Villigen, Switzerland\\
$^{7}$Institut Laue-Langevin, 6 rue Jules Horowitz, 38042 Grenoble, France\\}

\date{\today}
\begin{abstract}
We report a study of the basal-plane anisotropy of the superfluid density in underdoped YBa$_{2}$Cu$_{4}$O$_{8}$ (Y124), showing the effects of both the CuO$_2$ planes and the fully occupied CuO chains. From small-angle neutron scattering measurements of the vortex lattice, we can infer the superconducting (SC) properties for a temperature ($T$) range $T$~=~1.5 K to $T_{\textrm c}$ and magnetic induction $B$ from 0.1 to 6~T. We find that the superfluid density along {\textbf a} has a simple $d$-wave $T$-dependence. However, along {\textbf b} (the chain direction) the superfluid density falls much more rapidly with $T$ and also with increasing field. This strongly suggests the suppression of proximity-effect induced superconductivity in the CuO chains. In addition, our results do not support a common framework for the low field in-plane SC response in Y124 and related YBa$_{2}$Cu$_{3}$O$_{7}$, and also indicate that any magnetic field-induced charge-density-wave order in Y124 exists only for fields above 6~T.
\end{abstract}


\maketitle
\section{INTRODUCTION}
\label{sec:1Int}

The unifying structural constituent of all cuprate superconductors is the 2D CuO$_{2}$ planes. The YBaCuO family, YBa$_{2}$Cu$_{3}$O$_{7-\delta}$ (Y123) and YBa$_{2}$Cu$_{4}$O$_{8}$ (Y124), are special since they also host 1D CuO chains along the crystal \textbf{b}-axis. For structurally well-ordered chains the associated electronic states contribute to the Fermi surface. This makes YBaCuO a model system for studying low dimensional conductors in close proximity. Indeed, it has long been proposed that by proximity to the CuO$_{2}$ planes, the CuO chain states become superconducting (SC) below $T_{\rm c}$.~\citep{Atk95,Kre92} This is supported by the experimentally observed \emph{ab}-plane SC anisotropy where the London penetration depth is shorter for currents flowing along the \textbf{b}-axis than the \textbf{a}-axis. Moreover, the observed anisotropy is larger in Y124,~\citep{Bas95,Tal95} which displays two CuO chains per unit cell, than single-chained Y123.

While proximity-effect (PE) models provide an explanation for the in-plane SC anisotropy, a single framework for both Y123 and Y124 is not supported by experimental evidence. In Y123, a clear electronic anisotropy in the CuO$_{2}$ plane reported from transport~\citep{Dao10} and ARPES~\citep{Lu01} studies implies SC chain states are not required to explain the SC anisotropy. Measurements of the London penetration depth $\lambda$ give further information, since for a crystal axis $i$, $n_{i}(T)\propto\lambda_{i}^{-2}(T)$. In Y123, these show that the superfluid density $n$ along both \textbf{a} and \textbf{b} axes has a $d$-wave temperature- ($T$-) dependence,~\citep{Bon96,Car99,Whi11} which disagrees with the expectations of PE models.~\citep{Atk95,Xia96}

For Y124, the picture is somewhat unclear. It has been suggested that a positive low-$T$ curvature of both $n_{a}(T)$ and $n_{b}(T)$ observed by low field $\mu$SR is evidence for a two gap SC state in the CuO$_{2}$ plane.~\citep{Kha08} However, from other penetration depth measurements only $n_{b}(T)$ was observed to display a positive curvature, while $n_{a}(T)$ displayed $d$-wave behaviour.~\citep{Ser10} These results were argued as evidence for PE-induced SC chain states in Y124.

Here we present small-angle neutron scattering (SANS) measurements of the vortex lattice (VL) in Y124. Our measurements are conducted over a wide range of magnetic fields and temperatures, and the results cast important light on the plane-chain interplay in YBaCuO compounds. SANS experiments are a bulk probe of both the VL structure and the microscopic field distribution, each of which depend on the SC length-scales. Our single crystal samples of Y124 are naturally twin-free, and the stoichoimetric oxygen content makes the CuO chains effectively infinite in length. These properties suppress the vortex pinning effects seen in some Y123 samples, such as those due to twin-planes and oxygen vacancies.~\citep{Joh99,Bro04,Whi08} Therefore, the results of the present study on Y124 provide valuable comparison to the intrinsic VL properties only recently observed in twin-free and fully-oxygenated YBa$_{2}$Cu$_{3}$O$_{7}$ (Y1237).~\citep{Whi09,Whi11}

As will be seen in what follows, the VL in Y124 displays remarkably different properties to those observed in Y1237. This is apparent from measurements of both the VL structure presented in Sec.~\ref{sec:3Res_VL_Coord}, and the magnetic field- and $T$-dependence of the VL form factor presented in Sec.~\ref{sec:3VL_FF}. In Sec.~\ref{sec:4Discussion} we discuss the new insights provided by our results in connection with important topics relevant for both Y123 and Y124, such as proximity-effect induced CuO chain superconductivity, the basal plane superconducting anisotropy, and charge-density-wave order. Finally, a summary is presented in Sec.~\ref{sec:5Summ}. In Sec.~\ref{sec:2Exp} we begin by detailing the experimental method.

\section{EXPERIMENTAL METHOD}
\label{sec:2Exp}
Single crystals of YBa$_{2}$Cu$_{4}$O$_{8}$ were prepared as described in Ref.~\onlinecite{Kar99}. Each had approximate size 0.8 x 0.3 x 0.05~mm$^{3}$, and the longest side parallel to the \textbf{b}-axis. To obtain a sample mass suitable for neutron scattering experiments, 130 single crystals of total volume of 3.95 $\times$ 10$^{-5}$ m$^{2}$ were mounted onto a thin aluminum plate, each with their \textbf{c}-axis perpendicular to the plate, and \textbf{a}-axis vertical. The resulting mosaic had $T_{\rm c}\simeq79.9$~K with $\Delta T_{\rm c}\simeq2$~K.

The SANS experiments were conducted at the Institut Laue-Langevin, Grenoble, France, and the Swiss spallation neutron source (SINQ), Paul Scherrer Institut, Villigen, Switzerland. The sample mosaic was installed inside either a 6~T or 11~T horizontal field cryomagnet that provided a base temperature ($T$) of 1.5~K. The crystal \textbf{c}-axis was parallel to the applied magnetic field ($\mu_{0}H$), with both approximately parallel to the incident neutron beam. Cold neutrons ($\lambda_{n}=$ 0.6 to 1.66~nm, with FWHM spread of 10~\%) were collimated over 8-14~m, and after diffracting from the sample, were counted on a position-sensitive multidetector placed 8-14~m away. Measurements carried out at $T>T_{\rm c}$ were subtracted from those done at low $T$, in order to leave just the VL signal.

Due to the intrinsically twin-free and stoichiometric properties of the crystals, VL pinning is expected to be suppressed in our sample. Nonetheless, weak pinning due to residual crystal defects may still be expected. In these circumstances, cooling through $T_{\rm c}$ in a magnetic field that is weakly oscillating around the target field allows the vortices to overcome the pinning potential and attain a coordination closer to equilibrium. Therefore, all measurements reported in the paper were conducted on VLs prepared using a weakly oscillatory magnetic field component of $\pm$0.02 - 0.05~T during either cooling or warming to the target $T$. When at the intended $T$, the field was held stationary at the target value when conducting the SANS measurements.

%As reported in the main text, we found that even in spite of applying a weakly oscillating the field, the VL structure nevertheless became `frozen', or pinned, at $T\sim$30~K, thus indicating the irreversibility line in our sample.

\section{RESULTS}
\label{sec:3Res}

\subsection{Vortex Lattice Structure}
\label{sec:3Res_VL_Coord}

\begin{figure}
\includegraphics[width=0.48\textwidth]{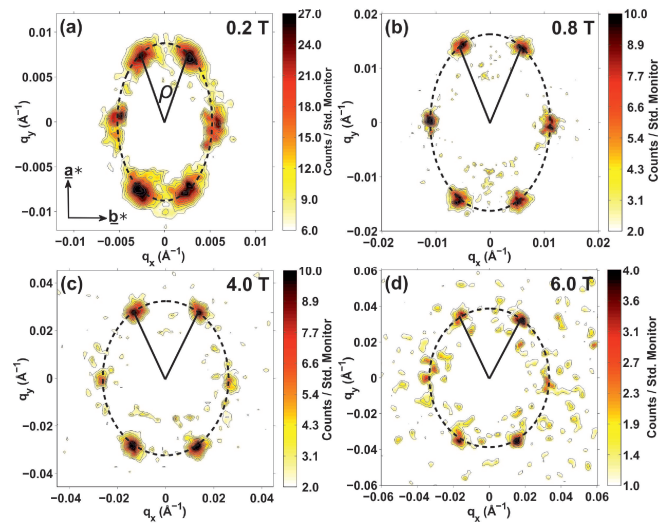}
\caption{(Color online) VL diffraction patterns obtained in Y124 at 1.5~K, and in $\mu_{0}H\parallel$~\textbf{c} of (a) 0.2~T, (b) 0.8~T, (c) 4.0~T and (d) 6.0~T. Each image is the sum of scattering from the VL as the sample is both tilted and rotated so that the Bragg condition is satisfied at the detector for the different diffraction spots. In each image, solid lines show the reciprocal VL basis vectors, and dashed line ellipses emphasise the VL anisotropy. The VL opening angle $\rho$ is defined in (a).}
\label{fig:FieldPics}
\end{figure}

\begin{figure}
\includegraphics[width=0.48\textwidth]{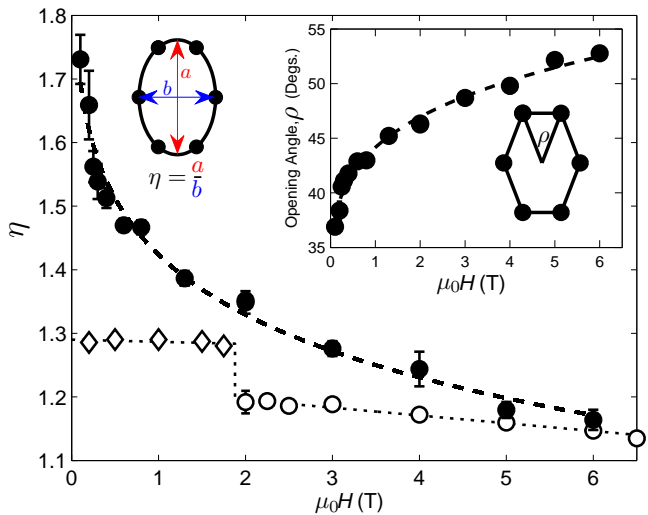}
\caption{(Color online) The $\mu_{0}H$-dependence of the VL distortion parameter $\eta$, defined by the inset sketch. Results for Y124 are shown with filled symbols. Empty symbols denote similar Y1237 data.~\citep{Whi09,Whi11} Inset: the $\mu_{0}H$-dependence of the VL opening angle $\rho$ in Y124. All lines are guides for the eye.}
\label{fig:structure}
\end{figure}

Fig.~\ref{fig:FieldPics} shows VL diffraction patterns obtained from Y124 at $T$=1.5~K, and over the observable field range up to $\mu_{0}H\parallel$~\textbf{c}~=~6.0~T. At all fields, the VL forms a single distorted hexagonal domain aligned with the crystal axes. We quantify the distortion in terms of the axial ratio of the ellipse that overlays the Bragg spots, $\eta$ which is related to the VL opening angle $\rho$ by $\eta=(\sqrt{3}\textrm{tan}\left(\rho/2\right))^{-1}$. The field-dependencies of both $\eta$ and $\rho$ at $T$=1.5~K are shown in Fig.~\ref{fig:structure}. With increasing $\mu_{0}H$, $\eta$ reduces and the VL structure becomes increasingly isotropic. For comparison, in Fig.~\ref{fig:structure} we include equivalent data for Y1237.~\citep{Whi09,Whi11} It is clear that the low field-dependence of $\eta$ is much stronger in Y124 than in Y1237. At higher fields however, the two compounds display more comparable behavior.

For high-$\kappa$ materials, and low $\mu_{0}H\parallel$~\textbf{c}, anisotropic local London theory~\citep{Kog81,Cam88,Thi89} gives the VL distortion parameter $\eta$ equal to the in-plane penetration depth anisotropy $\gamma_{ab}=\lambda_{a}/\lambda_{b}$. The sign of the observed VL distortion in Y124 shows that $\lambda_{a}>\lambda_{b}$, so the supercurrent density is larger along the CuO chain direction. Moreover, $\eta$ is larger in Y124 than Y1237 for the same fields, thus confirming the more anisotropic SC state in the double-chained compound.~\citep{Bas95} For both materials, the $\mu_{0}H$-induced suppression of $\eta$ implies a reduction in $\gamma_{ab}$ within local theory. However, care must taken if assuming $\eta=\gamma_{ab}$ across the entire $T$- and $\mu_{0}H$-range. In particular, at high $\mu_{0}H$ the equality becomes increasingly invalid due to nonlocal effects.~\citep{Whi11} Nonetheless, at low $\mu_{0}H$ where local theory is most valid, the two compounds display clearly different behavior. In Y1237 $\eta$ is constant up to a kink at a VL structure transition at $\sim$2~T, and after that falls monotonically with field.~\citep{Whi09} In contrast, $\eta$ in Y124 varies smoothly over the entire field range, and falls quickly at low fields. These observations show that even close to the local regime, the nonlocal interactions present in each compound lead to markedly different VL properties. In turn, this evidences the very different basal-plane SC responses for Y124 and Y1237.

Since local theory provides no constraint on the VL orientation for $\mu_{0}H\parallel$~\textbf{c}, even a weak additional interaction can give a preferred VL alignment. In anisotropic materials like YBaCuO, this can be due to nonlocal interactions between the VL and the system anisotropies such as those of the Fermi surface~\citep{Kog97a} and the SC gap.~\citep{Fra97,Ich99} Determining which is most influential requires first-principle numerical calculations that include the details of both anisotropies.~\citep{Nak02} These calculations will also shed light on why the single VL orientation observed in Y124 is seen only in an intermediate field range 2~T$<\mu_{0}H<$6.7~T in Y1237.~\citep{Whi09} Furthermore, in Y1237 a high-field transition, proposed to be driven by the $d$-wave gap,~\citep{Whi09} separates the intermediate field structure from a rhombic one that evolves smoothly to become almost square by 10.8~T. No sign of a similar transition is observed for $\mu_{0}H\leq6.0$~T in our Y124 sample, while for higher fields the SANS VL signal is too weak to observe. To search for a square-like VL structure in Y124, high field SANS studies on larger samples are required.

\subsection{Vortex Lattice Form Factor}
\label{sec:3VL_FF}

%The VL form factor, $F(\textbf{q})$ at a wavevector \textbf{q}, is the magnitude of the Fourier component $F(\textbf{q})$ that describes spatial variation of the internal field in the mixed state.

\begin{figure}
\includegraphics[width=0.48\textwidth]{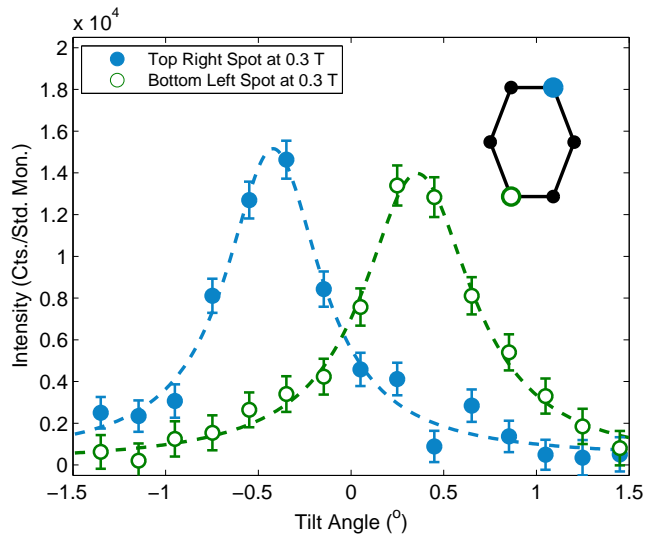}
\caption{Typical examples of the angular dependence of the diffracted intensity (rocking curves) obtained from the VL in Y124 at $\mu_{0}H=0.3$~T and $T=$1.5~K. Dashed lines correspond to a fit of a Lorentzian lineshape to each curve. The tilt angle corresponds to the rotation angle of the sample and cryomagnet around the horizontal axis. The inset sketch shows the VL structure, with two of the VL spots denoted by symbols that match those of the associated rocking curves.}
\label{fig:SI_fig2}
\end{figure}

Next we discuss measurements of the VL form factor, $F(\textbf{q})$ which is the Fourier transform of the magnetic field modulation in the mixed state. Experimentally, $F(\textbf{q})$  at the wavevector \textbf{q} is obtained from the integrated intensity $I_{\textbf{q}}$ of a VL diffraction spot as it is rotated through the Bragg condition at the detector. Fig.~\ref{fig:SI_fig2} shows the typical angular variation of the diffracted intensity (the rocking curve) measured from the low field VL in Y124. The quantity $I_{\textbf{q}}$ is obtained by integrating the area underneath the Lorentzian line shape used to fit the curve, and is related to $|F(\textbf{q})|$ by~\citep{Chr77}
\begin{equation}
I_{\textbf{q}}=2\pi\phi(\gamma/4)^{2}V\lambda_{n}^2\Phi_{0}^{-2}\textbf{q}^{-1}\left|F(\textbf{q})\right|^{2}.
\label{eq:Christen}
\end{equation}
Here $\phi$ is the intensity of the incident neutron beam, $\gamma$ is the neutron magnetic moment in nuclear magnetons, $\Phi_{0}$ is the flux quantum, $V$ is the sample volume, and $\lambda_{n}$ the neutron wavelength.

%$|F(\textbf{q})|^2$ is proportional to the integrated intensity $I_{\textbf{q}}$ of a VL diffraction spot as it is rotated through the Bragg condition (see Ref.~\onlinecite{Sup}).
\subsubsection{Field-dependence of the form factor}
\label{sec:3VL_FF_H}

\begin{figure}
\includegraphics[width=0.48\textwidth]{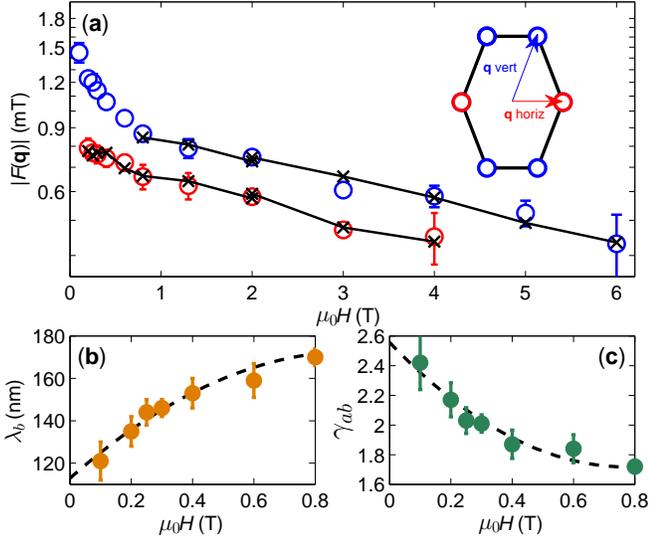}
\caption{(Color online) (a) A semilog plot of the $\mu_{0}H$-dependence of the VL form factors, $|F\left(\textrm{\textbf{q} vert}\right)|$ and $|F\left(\textrm{\textbf{q} horiz}\right)|$ at $T=$~1.5 K. The inset sketch of the VL structure defines the two types of Bragg spot, the form factors of which are treated separately. Black lines are fits of Eq.~\ref{eq:London} to the curves using a $\mu_{0}H$-independent parameter sets (see text for details). (b) and (c) respectively show the field-dependence of $\lambda_{b}$ and $\gamma_{ab}$, after varying $\lambda_{b}$ in Eq.~\ref{eq:London} to obtain a good description of the low field $|F\left(\textrm{\textbf{q} vert}\right)|$ data in (a) (see text for details). Dashed lines are guides for the eye.}
\label{fig:FF}
\end{figure}

Fig.~\ref{fig:FF}~(a) shows the $\mu_{0}H$-dependence at $T$=1.5~K of the VL form factor in Y124. In many strongly type-II superconductors, the observed fall off with $\mu_{0}H$ can be represented by the anisotropic London model extended by a gaussian cutoff which represents the finite size of the vortex cores.~\citep{Yao97,Whi11,Esk11} The cutoff leads to an expected exponential reduction in $F(\textbf{q})$ with $\mu_{0}H$. The model is valid for both $\kappa\gg1$ and $H\ll H_{\rm c2}$, and is in general is $T$-dependent:

%In many strongly type-II superconductors, the observed fall off with field can be represented by the anisotropic London model extended by a gaussian cutoff which represents the finite size of the vortex cores,~\citep{Yao97,Whi11,Esk11} and gives an exponential reduction in $F$ with field. The model is valid for $\kappa\gg1$, and in general is $T$-dependent:
\begin{eqnarray}
F\left(\textbf{q},T\right)=\frac{\langle B\rangle \textrm{exp}\left(-0.44\left(\textrm{q}_{x}^{2}\xi_{b}(T)^{2}+\textrm{q}_{y}^{2}\xi_{a}(T)^{2}\right)\right)}{1+(\textrm{q}_{x}^{2}\lambda_{a}(T)^{2}+\textrm{q}_{y}^{2}\lambda_{b}(T)^{2})},
\label{eq:London}
\end{eqnarray}
where $\langle B\rangle$ is the internal induction. $\xi_{i}(T)$ and $\lambda_{i}(T)$ respectively denote the GL coherence lengths and London penetration depths along directions $i$. $\textrm{q}_{x}$ and $\textrm{q}_{y}$ denote components of $\textbf{q}$ parallel to \textbf{b}$^{\ast}$ and \textbf{a}$^{\ast}$, respectively. For all fits we used the experimentally observed $\textbf{q}$-values.

We note that the horizontal $|F\left(\textrm{\textbf{q} horiz}\right)|$ spots, which have $\textbf{q}\parallel\textbf{b}^{\ast}$, lie close to a straight line in the semilog plot in Fig.~\ref{fig:FF}~(a) and so can be fitted with constant values of just two parameters $\xi_{\rm b}$ and $\lambda_{\rm a}$ in Eq.~\ref{eq:London}. However, the top/bottom $|F\left(\textrm{\textbf{q} vert}\right)|$ spots have an anomalous behaviour for $\mu_{0}H<$~0.8~T. Concentrating therefore on fitting just the $|F\left(\textrm{\textbf{q} horiz}\right)|$ spots, we obtain $\xi_{b}$=~3.6(2)~nm and $\lambda_{a}$=~293(3)~nm at 1.5~K. The value of $\xi_{b}$ implies $H_{c2}$$\sim$25~T, which is below the recently reported value of 44~T,~\cite{Cha12c,Ram12} and suggests a contribution of weak VL disorder to the variation of the form factor.~\citep{Whi11}

Extending the analysis to the $|F(\textrm{\textbf{q} vert})|$ spots, it is clear that above 0.8~T the model can be applied to fit them too, and we obtain in addition $\lambda_{b}$=~170(4)~nm and $\xi_{a}$=~3.7(2)~nm. To capture the behaviour of $|F(\textrm{\textbf{q} vert})|$ for $\mu_{0}H<$~0.8~T, we first note that according to Eq.~\ref{eq:London}, $\xi_{i}$ has little influence at low fields. Therefore, to describe the data $\lambda_{b}$ must become smaller at low field, which corresponds to an increase in the superfluid density for currents along \textbf{b}. Fig.~\ref{fig:FF}~(b) shows the low $\mu_{0}H$-dependence of $\lambda_{b}$ which, for all other parameters $\mu_{0}H$-independent, gives calculated $|F(\textrm{\textbf{q} vert})|$ values consistent with the experimental data. Using the $\mu_{0}H$-dependent values of $\lambda_{b}$, and $\lambda_{a}$=~293(3)~nm, the low $\mu_{0}H$-dependence of $\gamma_{ab}$ at 1.5~K is shown in Fig.~\ref{fig:FF}~(c).

From our analysis of the $\mu_{0}H$-dependent form factor at 1.5~K, there are apparently two disagreements with anisotropic London theory: a) $\gamma_{ab}$ is always $>\eta$ at the same field (compare Fig.~\ref{fig:FF}~(c) with Fig.~\ref{fig:structure}), and b) the form factors for the two spot types do not become equal at low fields.~\citep{Whi11} Both these discrepancies may arise if the VL structure is $T$-dependent, and becomes pinned so that the value of $\eta$ at 1.5~K does not represent the true SC anisotropy, which is $\gamma_{ab}$. To demonstrate that this is the case, we consider low field results at higher $T$.

\subsubsection{Temperature-dependence of the form factor}
\label{sec:3VL_FF_T}

$T$-dependent measurements of the VL form factor at low field provide direct insight concerning both the anisotropy of the superfluid density and the underlying gap structure. The intensive nature of these measurements means that there was insufficient neutron beamtime to record full rocking curves, and hence $I_{\textbf{q}}$, at each $T$. Therefore, measurements were done just at the Bragg angle (at the peak of the rocking curve), with full rocking curve measurements done at selected $T$s to confirm the $T$-independence of the rocking curve width. All $T$-dependent measurements were done by warming scans conducted after an initial oscillation field-cool to 1.5~K.

\begin{figure}
\centering
\includegraphics[width=0.48\textwidth]{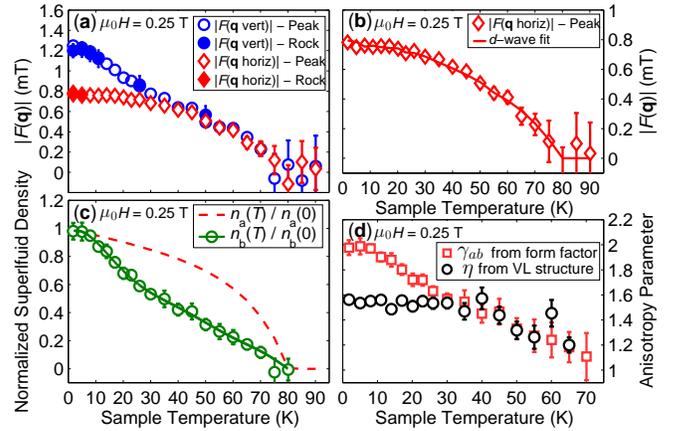}
\caption{(color online) (a) A warming $T$-dependence at $\mu_{0}H=0.25$~T of the VL form factors. Filled and empty symbols respectively denote full rocking curve, and rocking curve peak measurements. (b) The $T$-dependence of $|F\left(\textrm{\textbf{q} horiz}\right)|$. The line is a $d$-wave model fit to the data with $\Delta_{0}(0)$=~24(2)~meV. (c) Normalized superfluid densities $n_{a}(T)$ and $n_{b}(T)$. The values for $n_{b}$($T$) are extracted as described in the text. The solid line is a guide to the eye. The dashed line is the ideal curve for $n_{a}(T)$ obtained from the fit in panel (b). (d) The $T$-dependences at $\mu_{0}H=0.25$~T of the anisotropy parameters $\gamma_{ab}$ and $\eta$.
\label{fig:multipanel_tdep}}
\end{figure}

Fig.~\ref{fig:multipanel_tdep}~(a) shows the $T$-dependence of the form factors at $\mu_{0}H=0.25$~T. We see that on warming the base $T$ form factor anisotropy is suppressed, so that the form factors eventually become equal as expected within London theory. To show how both $\gamma_{ab}$ and $\eta$ compare at higher $T$, we firstly need to calculate $\lambda_{a}(T)$ from the $|F\left(\textrm{\textbf{q} horiz}\right)|$ data, and then subsequently we can extract $\lambda_{b}(T)$ from the $|F\left(\textrm{\textbf{q} vert}\right)|$ data. For the first step, we see from Eq.~\ref{eq:London} that $\textrm{q}_{y}$=~0 for the $|F\left(\textrm{\textbf{q} horiz}\right)|$ spots, and so they are sensitive only to the $T$ variation of $\xi_{b}(T)$ and $\lambda_{a}(T)$ from their previously established base $T$ values. The $T$ variations of both $\xi_{b}(T)$ and $\lambda_{a}(T)$ are calculated following the same approach reported in Ref.~\onlinecite{Whi11}. For calculating $\lambda_{a}(T)$, we compute the $T$-dependent quasiparticle spectrum expected over a standard $d$-wave SC gap on a single quasi-cylindrical sheet, and for which the zero $T$ gap magnitude, $\Delta_{0}(T=0)$ is a free parameter.~\citep{Not3}

%The value of $\lambda_{a}(0)$ agrees well with that determined from the analysis of the $\mu_{0}H$-dependent form factor at base $T$,

In principle therefore, the fit of the $|F\left(\textrm{\textbf{q} horiz}\right)|$ data is dependent on three parameters, $\xi_{b}(0)$, $\lambda_{a}(0)$ and $\Delta_{0}(0)$. We found that the fit was insensitive to $\xi_{b}(0)$, and so this parameter was fixed at 3.6~nm as determined in Sec.~\ref{sec:3VL_FF_H}. As shown in Fig.~\ref{fig:multipanel_tdep}~(b), the $T$-dependence of $|F\left(\textrm{\textbf{q} horiz}\right)|$ is well described by the $d$-wave model for the superfluid density, and the two remaining free parameters are fitted to be $\lambda_{a}$=~290(3)~nm and $\Delta_{0}(0)$=~24(2)~meV. The good fit and agreement of the parameter values with those reported elsewhere~\citep{Bas95,Sok95,Kha08} confirms $n_{a}(T)$ to be controlled by a single $d$-wave SC gap.

Next we extract the $T$-dependence of $n_{b}(T)$. By using the $|F\left(\textrm{\textbf{q} vert}\right)|$ data shown in Fig.~\ref{fig:multipanel_tdep}~(a), the $d$-wave model fit for $n_{a}(T)$, and the zero-$T$ values for $\xi_{i}$ obtained in Sec.~\ref{sec:3VL_FF_H}, we can solve for the only remaining unknown in Eq.~\ref{eq:London} which is $\lambda_{b}(T)(\propto 1/\sqrt{n_{b}(T)})$. The extracted $T$-dependence of $n_{b}(T)$ is shown in Fig.~\ref{fig:multipanel_tdep}~(c). The positive curvature observed below $\sim T_{\rm c}/2$ is inconsistent with usual $d$-wave behaviour, and instead evidences a multi-gap SC response along \textbf{b}. We also note that the extrapolated value $\lambda_{b}(0)$=~145(2)~nm at this field agrees with that shown in Fig.~\ref{fig:FF}~(b).

Fig.~\ref{fig:multipanel_tdep}~(d) shows the $T$-dependence of the SC anisotropy parameters at $\mu_{0}H=0.25$~T. Here $\gamma_{ab}$ is determined at each $T$ using the absolute $n_{i}(T)$ data, while $\eta$ is obtained directly from the VL structure. Above $\sim$30~K, $\gamma_{ab}$ and $\eta$ agree well, as expected within anisotropic London theory. For $T<30$~K however, a clear difference between $\gamma_{ab}$ and $\eta$ emerges, as $\eta$ varies only weakly on cooling, while $\gamma_{ab}$ increases smoothly. This behavior, and that of Fig.~\ref{fig:multipanel_tdep}~(a), is explained most simply if the VL structure becomes frozen on cooling below $\sim$30~K and is thus unable to evolve further on cooling. Importantly then, the \emph{intrinsic} low $T$ SC anisotropy is \emph{only} given by $\gamma_{ab}$ as evaluated using $\lambda_{i}$ values obtained from analysing the form factor. From our $T$-dependent data, we find that at $0.25$~T, $\gamma_{ab}$=1.97(4) by 1.5~K, and from data at $0.4$~T (see Appendix~\ref{sec:6AppA}), $\gamma_{ab}$=1.80(2). These values agree with those obtained independently from the field-dependent form factor analysis [Fig~\ref{fig:FF}~(c)]. Also from Fig.~\ref{fig:FF}~(c), our extrapolation to zero field of $\gamma_{ab}$=2.57(5) agrees well with the value of 2.5 determined by far infra-red spectroscopy.~\citep{Bas95}

\section{DISCUSSION}
\label{sec:4Discussion}
Our analysis of the $T$-dependent form factor at 0.25~T shows $n_{a}(T)$ to be mainly sensitive to a single $d$-wave SC gap, while $n_{b}(T)$ requires a multi-gap description. Evidence for a contribution to the multi-gap response of $n_{b}(T)$ is provided by ARPES where a $\sim$5~meV gap is observed directly at the crossing point between plane and chain Fermi surfaces.~\citep{Kon10} This ties the origin of the gap strongly to plane-chain hybridization as expected within the PE models.~\citep{Atk95,Xia96} The field-dependence of $\lambda_{b}$ at 1.5~K [Fig.~\ref{fig:FF}~(c)] likely reflects the quenching of the 5~meV gap, or an as yet unobserved part of the gap structure. We also expect that the sharp change in the $\mu_{0}H$-dependence of the form factor ratio observed at $\sim$0.8~T [Fig.~\ref{fig:FF}~(a)] marks the field-scale of this quenching, such as a critical field of the PE-induced SC response.

Next we compare the low $\mu_{0}H$ SC responses of Y124 and Y1237. In Y124, the $T$-dependent forms for both $n_{a}(T)$ and $n_{b}(T)$ at $\mu_{0}H=0.25$~T [Fig.~\ref{fig:multipanel_tdep}~(c)] are qualitatively similar to those reported at very low field in Ref.~\onlinecite{Ser10}, and are consistent with a PE model where the plane-chain coupling is mediated via single electron tunnelling.~\citep{Xia96} However, this model can not explain low field SANS data collected on Y1237, where both $n_{a}(T)$ and $n_{b}(T)$ display $d$-wave-like $T$-dependencies.~\citep{Whi11,Not1} Moreover, the direct observation of a PE-induced SC gap analogous to that seen in Y124~\citep{Kon10} is not reported for Y1237. To explain this suprising difference between the two materials, a possibility is that the chain states in Y1237 are non-SC and the SC anisotropy is intrinsic to the planes. On the other hand, a $d$-wave behavior for each of $n_{a}(T)$ and $n_{b}(T)$ is consistent with calculations that consider \emph{intrinsically} SC chains coupled to the planes by a Josephson-type pair tunnelling.~\citep{Xia96} If the latter is true, the $d$-wave $T$-dependence of $\lambda_{b}$ in Y1237 implies that there is a node in the SC gap on the chain FS. Both new ARPES experiments and detailed calculations can shed light on these proposals.

For $\mu_{0}H>2$~T, the $\mu_{0}H$-dependence of the low-$T$ VL properties in Y124 and Y1237~\citep{Whi09,Whi11} is more comparable, and indicates the two compounds display high-field SC regimes that become more similar. In both materials $\gamma_{ab}$ is always $>1$ which, to a first approximation, shows the persistence of the in-plane SC anisotropy to high field. Whether this anisotropy reflects a persistent contribution due to SC chain states, an intrinsic CuO$_{2}$ plane anisotropy, or is even related to the Fermi surface reconstruction in these materials,~\citep{Doi07,Yel08,Ban08} remains an important open question.

Finally, we comment on an implication of our study concerning the interplay between co-existing SC and charge-density-wave (CDW) orders observed in a range of underdoped Y123 samples.~\citep{Ghi12,Cha12b,Ach12} The CDW and SC orders compete since when applying a magnetic field to suppress superconductivity, the CDW order is observed to grow.~\citep{Cha12b} In Y124, no CDW order has yet been observed at zero field. Nevertheless, underdoped Y123 and Y124 both display comparable quantum oscillations frequencies,~\citep{Doi07,Yel08,Ban08} and negative values of the Hall coefficient at low $T$.~\citep{LeB07,Rou10} This indicates similar $\mu_{0}H$-driven reconstructions of the Fermi surface to occur in both Y123 and Y124, and which all likely involve CDW order. We speculate that the effect on the VL due to $\mu_{0}H$-induced CDW order in Y124 may be similar to that due to $\mu_{0}H$-induced spin-density-wave (SDW) order in La$_{2-x}$Sr$_{x}$CuO$_{4}$, $x=0.145$.~\citep{Cha12} There it was reported that the slope of the $\mu_{0}H$-dependent VL form factor increases sharply at the onset of SDW order, after which the form factor falls with $\mu_{0}H$ more rapidly than describable using conventional models.~\citep{Yao97,Cle75} The increase in slope was explained as caused by a disordering of the VL, which evidences the competition between the SC and SDW orders.~\citep{Cha12} Since in Fig.~\ref{fig:FF}~(a) we observe no sharp change in slope of the form factor beyond that more easily understood in terms of a quenching of the PE, our results appear to limit any $\mu_{0}H$-induced CDW order in Y124 to fields $>6$~T.

\section{SUMMARY}
\label{sec:5Summ}
In summary we have studied the VL in YBa$_{2}$Cu$_{4}$O$_{8}$ (Y124) with $\mu_{0}H\parallel$~\textbf{c}. At all fields, the VL structure is distorted due to the in-plane SC anisotropy. The VL distortion is suppressed with increasing $\mu_{0}H$ which most likely reflects a quenching of a proximity-effect-induced SC gap involving chain states. Our results rule out a common framework for the low field in-plane SC response of Y124 and YBa$_{2}$Cu$_{3}$O$_{7}$, and also indicate any $\mu_{0}H$-induced CDW order in Y124 exists only for $\mu_{0}H>6$~T.

\begin{center}
\textbf{ACKNOWLEDGEMENTS}
\end{center}
We acknowledge discussions with J.~Chang, A.T.~Holmes, M.~Ichioka and K.~Machida. SANS experiments were performed at the ILL, Grenoble, France, and the Swiss spallation neutron source, SINQ, PSI, Switzerland. We acknowledge financial support from the EPSRC of the UK, the University of Birmingham, the Swiss NCCR and its program MaNEP, and from the European Commission under the 6$^{th}$ Framework Programme though the Key Action: Strengthening the European Research Area, Research Infrastructures, Contract No. RII3-CT-2003-505925.

\appendix
\section{Temperature-dependent form factor at $\mu_{0}H=0.4$~T}
\label{sec:6AppA}
\begin{figure}
\includegraphics[width=0.48\textwidth]{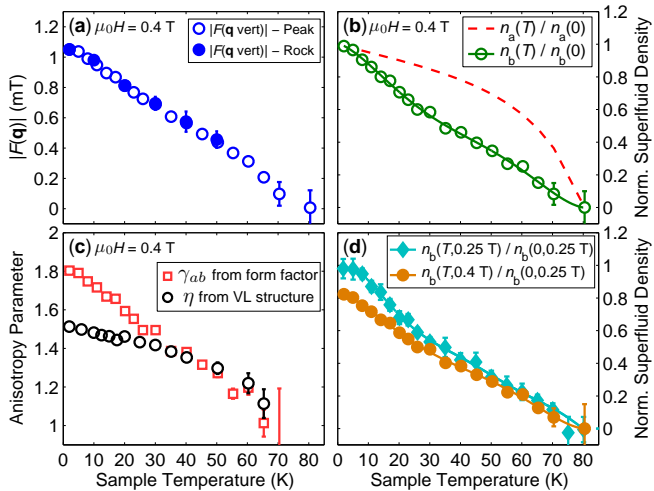}
\caption{(a) A warming $T$-dependence at $\mu_{0}H=0.4$~T of the $|F\left(\textrm{\textbf{q} vert},T\right)|$ VL form factors. Filled and empty symbols respectively denote measurements of full rocking curves, and measurements done just at the rocking curve peak. (b) Normalised superfluid densities along \textbf{a} ($n_{a}(T)$) and \textbf{b} ($n_{b}(T)$). The dashed line is the ideal curve for $n_{a}(T)$ obtained from the fit presented in Fig.~\ref{fig:multipanel_tdep}(b) of the main paper. The circles are values of $n_{b}$($T$) extracted as explained in the text, and the solid line is a guide to the eye. (c) The $T$-dependences at $\mu_{0}H=0.4$~T of $\gamma_{ab}$ determined from the analysis of the $T$-dependent form factor, and $\eta$ determined from $T$-dependent measurements of the VL structure. (d) The $T$-dependence of $n_{b}$($T$) extracted at both $\mu_{0}H=0.25$~T and $\mu_{0}H=0.4$~T. The data at $\mu_{0}H=0.4$~T have been normalized using the extrapolated zero-$T$ value of $n_{b}(0)$ at $\mu_{0}H=0.25$~T.}
\label{fig:SI_fig1}
\end{figure}

Here we describe the analysis of $T$-dependent VL form factor data measured at $\mu_{0}H=0.4$~T. Unlike the data recorded at $\mu_{0}H=0.25$~T where the $T$-dependence of both $|F\left(\textrm{\textbf{q} vert},T\right)|$ and $|F\left(\textrm{\textbf{q} horiz},T\right)|$ form factors was measured, at $\mu_{0}H=0.4$~T we only recorded the former. The $T$-dependent $|F\left(\textrm{\textbf{q} vert},T\right)|$ data measured at $\mu_{0}H=0.4$~T are shown in Fig.~\ref{fig:SI_fig1}(a).

Next we determine the $T$-dependence of $n_{b}(T)$ at 0.4~T from the data shown in Fig.~\ref{fig:SI_fig1}(a). We follow the same approach used when analyzing the $\mu_{0}H=0.25$~T data [Sec.~\ref{sec:3VL_FF_T}], where we solve Eq.~\ref{eq:London} at each $T$ to find $\lambda_{b}(T)(\propto 1/\sqrt{n_{b}(T)})$. In this instance, the calculation of $\lambda_{a}(T)$ is done using the values $\lambda_{a}(0)=290$~nm, $\Delta_{0}(0)=24$~meV determined from the fit of the $|F\left(\textrm{\textbf{q} horiz},T\right)|$ data shown in Fig.~\ref{fig:multipanel_tdep}(b). For the zero $T$ coherence lengths, we again assume that $\xi_{a}(0)=3.7$~nm and $\xi_{b}(0)=3.6$~nm as reported in Sec.~\ref{sec:3VL_FF_H}. The extracted $T$-dependence of $n_{b}(T)$ at $\mu_{0}H=0.4$~T is shown in Fig.~\ref{fig:SI_fig1}(b). Similarly as seen at $\mu_{0}H=0.25$~T, $n_{b}(T)$ at 0.4~T displays a positive curvature in the low $T$ region which is incompatible with a simple $d$-wave superconducting (SC) gap function.

Using the both the extracted $T$-dependence of $n_{b}(T)$, and the assumed form of $n_{a}(T)$, in Fig.~\ref{fig:SI_fig1}(c) we plot the $T$-dependence of the anisotropy parameter $\gamma_{ab}$. For comparison, we also show the $T$-dependence of $\eta$ determined from direct measurements of the $T$-dependent VL structure. Again similarly as seen at $\mu_{0}H=0.25$~T, a clear disagreement between $\gamma_{ab}$ and $\eta$ emerges for $T\lesssim30$~K thus marking the irreversibility $T$ at this field. The values of $\gamma_{ab}$ at low $T$ are thus larger than would be deduced solely by equating $\gamma_{ab}=\eta$ as expected in the local London approximation, though comparatively smaller than at $\mu_{0}H=0.25$~T. This reflects the $\mu_{0}H$-induced suppression of $n_{b}(T)$, which is shown in Fig.~\ref{fig:SI_fig1}(d). Here we make a relative comparison between the $n_{b}(T)$ curves extracted at both $\mu_{0}H=0.25$~T and $\mu_{0}H=0.4$~T. For making this comparison, both curves are normalised to the extrapolated value of $n_{b}(0)$ at $\mu_{0}H=0.25$~T. This shows clearly how $n_{b}(T)$ is suppressed with $\mu_{0}H$, which we expect to reflect mostly a suppression of the proximity-effect induced SC response involving chain electron states.

\bibliography{YBCO_124_2}

%merlin.mbs apsrev4-1.bst 2010-07-25 4.21a (PWD, AO, DPC) hacked
%Control: key (0)
%Control: author (8) initials jnrlst
%Control: editor formatted (1) identically to author
%Control: production of article title (-1) disabled
%Control: page (0) single
%Control: year (1) truncated
%Control: production of eprint (0) enabled
\begin{thebibliography}{44}%
\makeatletter
\providecommand \@ifxundefined [1]{%
 \@ifx{#1\undefined}
}%
\providecommand \@ifnum [1]{%
 \ifnum #1\expandafter \@firstoftwo
 \else \expandafter \@secondoftwo
 \fi
}%
\providecommand \@ifx [1]{%
 \ifx #1\expandafter \@firstoftwo
 \else \expandafter \@secondoftwo
 \fi
}%
\providecommand \natexlab [1]{#1}%
\providecommand \enquote  [1]{``#1''}%
\providecommand \bibnamefont  [1]{#1}%
\providecommand \bibfnamefont [1]{#1}%
\providecommand \citenamefont [1]{#1}%
\providecommand \href@noop [0]{\@secondoftwo}%
\providecommand \href [0]{\begingroup \@sanitize@url \@href}%
\providecommand \@href[1]{\@@startlink{#1}\@@href}%
\providecommand \@@href[1]{\endgroup#1\@@endlink}%
\providecommand \@sanitize@url [0]{\catcode `\\12\catcode `\$12\catcode
  `\&12\catcode `\#12\catcode `\^12\catcode `\_12\catcode `\%12\relax}%
\providecommand \@@startlink[1]{}%
\providecommand \@@endlink[0]{}%
\providecommand \url  [0]{\begingroup\@sanitize@url \@url }%
\providecommand \@url [1]{\endgroup\@href {#1}{\urlprefix }}%
\providecommand \urlprefix  [0]{URL }%
\providecommand \Eprint [0]{\href }%
\providecommand \doibase [0]{http://dx.doi.org/}%
\providecommand \selectlanguage [0]{\@gobble}%
\providecommand \bibinfo  [0]{\@secondoftwo}%
\providecommand \bibfield  [0]{\@secondoftwo}%
\providecommand \translation [1]{[#1]}%
\providecommand \BibitemOpen [0]{}%
\providecommand \bibitemStop [0]{}%
\providecommand \bibitemNoStop [0]{.\EOS\space}%
\providecommand \EOS [0]{\spacefactor3000\relax}%
\providecommand \BibitemShut  [1]{\csname bibitem#1\endcsname}%
\let\auto@bib@innerbib\@empty
%</preamble>
\bibitem [{\citenamefont {Atkinson}\ and\ \citenamefont
  {Carbotte}(1995)}]{Atk95}%
  \BibitemOpen
  \bibfield  {author} {\bibinfo {author} {\bibfnamefont {W.~A.}\ \bibnamefont
  {Atkinson}}\ and\ \bibinfo {author} {\bibfnamefont {J.~P.}\ \bibnamefont
  {Carbotte}},\ }\href@noop {} {\bibfield  {journal} {\bibinfo  {journal}
  {Phys. Rev. B}\ }\textbf {\bibinfo {volume} {52}},\ \bibinfo {pages} {10601}
  (\bibinfo {year} {1995})}\BibitemShut {NoStop}%
\bibitem [{\citenamefont {Krezin}\ and\ \citenamefont {Wolf}(1992)}]{Kre92}%
  \BibitemOpen
  \bibfield  {author} {\bibinfo {author} {\bibfnamefont {V.~Z.}\ \bibnamefont
  {Krezin}}\ and\ \bibinfo {author} {\bibfnamefont {S.~A.}\ \bibnamefont
  {Wolf}},\ }\href@noop {} {\bibfield  {journal} {\bibinfo  {journal} {Phys.
  Rev. B}\ }\textbf {\bibinfo {volume} {46}},\ \bibinfo {pages} {6458}
  (\bibinfo {year} {1992})}\BibitemShut {NoStop}%
\bibitem [{\citenamefont {Basov}\ \emph {et~al.}(1995)\citenamefont {Basov},
  \citenamefont {Liang}, \citenamefont {Bonn}, \citenamefont {Hardy},
  \citenamefont {Dabrowski}, \citenamefont {Quijada}, \citenamefont {Tanner},
  \citenamefont {Rice}, \citenamefont {Ginsberg},\ and\ \citenamefont
  {Timusk}}]{Bas95}%
  \BibitemOpen
  \bibfield  {author} {\bibinfo {author} {\bibfnamefont {D.~N.}\ \bibnamefont
  {Basov}}, \bibinfo {author} {\bibfnamefont {R.}~\bibnamefont {Liang}},
  \bibinfo {author} {\bibfnamefont {D.~A.}\ \bibnamefont {Bonn}}, \bibinfo
  {author} {\bibfnamefont {W.~N.}\ \bibnamefont {Hardy}}, \bibinfo {author}
  {\bibfnamefont {B.}~\bibnamefont {Dabrowski}}, \bibinfo {author}
  {\bibfnamefont {M.}~\bibnamefont {Quijada}}, \bibinfo {author} {\bibfnamefont
  {D.~B.}\ \bibnamefont {Tanner}}, \bibinfo {author} {\bibfnamefont {J.~P.}\
  \bibnamefont {Rice}}, \bibinfo {author} {\bibfnamefont {D.~M.}\ \bibnamefont
  {Ginsberg}}, \ and\ \bibinfo {author} {\bibfnamefont {T.}~\bibnamefont
  {Timusk}},\ }\href@noop {} {\bibfield  {journal} {\bibinfo  {journal} {Phys.
  Rev. Lett.}\ }\textbf {\bibinfo {volume} {74}},\ \bibinfo {pages} {598}
  (\bibinfo {year} {1995})}\BibitemShut {NoStop}%
\bibitem [{\citenamefont {Tallon}\ \emph {et~al.}(1995)\citenamefont {Tallon},
  \citenamefont {Bernhard}, \citenamefont {Binninger}, \citenamefont {Hofer},
  \citenamefont {Williams}, \citenamefont {Ansaldo}, \citenamefont {Budnick},\
  and\ \citenamefont {Niedermayer}}]{Tal95}%
  \BibitemOpen
  \bibfield  {author} {\bibinfo {author} {\bibfnamefont {J.~L.}\ \bibnamefont
  {Tallon}}, \bibinfo {author} {\bibfnamefont {C.}~\bibnamefont {Bernhard}},
  \bibinfo {author} {\bibfnamefont {U.}~\bibnamefont {Binninger}}, \bibinfo
  {author} {\bibfnamefont {A.}~\bibnamefont {Hofer}}, \bibinfo {author}
  {\bibfnamefont {G.~V.~M.}\ \bibnamefont {Williams}}, \bibinfo {author}
  {\bibfnamefont {E.~J.}\ \bibnamefont {Ansaldo}}, \bibinfo {author}
  {\bibfnamefont {J.~I.}\ \bibnamefont {Budnick}}, \ and\ \bibinfo {author}
  {\bibfnamefont {C.}~\bibnamefont {Niedermayer}},\ }\href@noop {} {\bibfield
  {journal} {\bibinfo  {journal} {Phys. Rev. Lett.}\ }\textbf {\bibinfo
  {volume} {74}},\ \bibinfo {pages} {1008} (\bibinfo {year}
  {1995})}\BibitemShut {NoStop}%
\bibitem [{\citenamefont {Daou}\ \emph {et~al.}(2010)\citenamefont {Daou},
  \citenamefont {Chang}, \citenamefont {Leboeuf}, \citenamefont
  {Cyr-Choinière}, \citenamefont {Laliberté}, \citenamefont {Doiron-Leyraud},
  \citenamefont {Ramshaw}, \citenamefont {Liang}, \citenamefont {Bonn},
  \citenamefont {Hardy},\ and\ \citenamefont {Taillefer}}]{Dao10}%
  \BibitemOpen
  \bibfield  {author} {\bibinfo {author} {\bibfnamefont {R.}~\bibnamefont
  {Daou}}, \bibinfo {author} {\bibfnamefont {J.}~\bibnamefont {Chang}},
  \bibinfo {author} {\bibfnamefont {D.}~\bibnamefont {Leboeuf}}, \bibinfo
  {author} {\bibfnamefont {O.}~\bibnamefont {Cyr-Choinière}}, \bibinfo
  {author} {\bibfnamefont {F.}~\bibnamefont {Laliberté}}, \bibinfo {author}
  {\bibfnamefont {N.}~\bibnamefont {Doiron-Leyraud}}, \bibinfo {author}
  {\bibfnamefont {B.}~\bibnamefont {Ramshaw}}, \bibinfo {author} {\bibfnamefont
  {R.}~\bibnamefont {Liang}}, \bibinfo {author} {\bibfnamefont {D.~A.}\
  \bibnamefont {Bonn}}, \bibinfo {author} {\bibfnamefont {W.}~\bibnamefont
  {Hardy}}, \ and\ \bibinfo {author} {\bibfnamefont {L.}~\bibnamefont
  {Taillefer}},\ }\href@noop {} {\bibfield  {journal} {\bibinfo  {journal}
  {Nature}\ }\textbf {\bibinfo {volume} {463}},\ \bibinfo {pages} {519}
  (\bibinfo {year} {2010})}\BibitemShut {NoStop}%
\bibitem [{\citenamefont {Lu}\ \emph {et~al.}(2001)\citenamefont {Lu},
  \citenamefont {Feng}, \citenamefont {Armitage}, \citenamefont {Shen},
  \citenamefont {Damascelli}, \citenamefont {Kim}, \citenamefont {Ronning},
  \citenamefont {Shen}, \citenamefont {Bonn}, \citenamefont {Liang},
  \citenamefont {Hardy}, \citenamefont {Rykov},\ and\ \citenamefont
  {Tajima}}]{Lu01}%
  \BibitemOpen
  \bibfield  {author} {\bibinfo {author} {\bibfnamefont {D.~H.}\ \bibnamefont
  {Lu}}, \bibinfo {author} {\bibfnamefont {D.~L.}\ \bibnamefont {Feng}},
  \bibinfo {author} {\bibfnamefont {N.~P.}\ \bibnamefont {Armitage}}, \bibinfo
  {author} {\bibfnamefont {K.~M.}\ \bibnamefont {Shen}}, \bibinfo {author}
  {\bibfnamefont {A.}~\bibnamefont {Damascelli}}, \bibinfo {author}
  {\bibfnamefont {C.}~\bibnamefont {Kim}}, \bibinfo {author} {\bibfnamefont
  {F.}~\bibnamefont {Ronning}}, \bibinfo {author} {\bibfnamefont {Z.-X.}\
  \bibnamefont {Shen}}, \bibinfo {author} {\bibfnamefont {D.~A.}\ \bibnamefont
  {Bonn}}, \bibinfo {author} {\bibfnamefont {R.}~\bibnamefont {Liang}},
  \bibinfo {author} {\bibfnamefont {W.~N.}\ \bibnamefont {Hardy}}, \bibinfo
  {author} {\bibfnamefont {A.~I.}\ \bibnamefont {Rykov}}, \ and\ \bibinfo
  {author} {\bibfnamefont {S.}~\bibnamefont {Tajima}},\ }\href@noop {}
  {\bibfield  {journal} {\bibinfo  {journal} {Phys. Rev. Lett.}\ }\textbf
  {\bibinfo {volume} {86}},\ \bibinfo {pages} {4370} (\bibinfo {year}
  {2001})}\BibitemShut {NoStop}%
\bibitem [{\citenamefont {Bonn}\ \emph {et~al.}(1996)\citenamefont {Bonn},
  \citenamefont {Kamal}, \citenamefont {Bonakdarpour}, \citenamefont {Ruixing},
  \citenamefont {Hardy}, \citenamefont {Homes}, \citenamefont {Basov},\ and\
  \citenamefont {Timusk}}]{Bon96}%
  \BibitemOpen
  \bibfield  {author} {\bibinfo {author} {\bibfnamefont {D.~A.}\ \bibnamefont
  {Bonn}}, \bibinfo {author} {\bibfnamefont {S.}~\bibnamefont {Kamal}},
  \bibinfo {author} {\bibfnamefont {A.}~\bibnamefont {Bonakdarpour}}, \bibinfo
  {author} {\bibfnamefont {R.}~\bibnamefont {Ruixing}}, \bibinfo {author}
  {\bibfnamefont {W.~N.}\ \bibnamefont {Hardy}}, \bibinfo {author}
  {\bibfnamefont {C.~C.}\ \bibnamefont {Homes}}, \bibinfo {author}
  {\bibfnamefont {D.~N.}\ \bibnamefont {Basov}}, \ and\ \bibinfo {author}
  {\bibfnamefont {T.}~\bibnamefont {Timusk}},\ }\href@noop {} {\bibfield
  {journal} {\bibinfo  {journal} {Czech. J. Phys.}\ }\textbf {\bibinfo {volume}
  {46}},\ \bibinfo {pages} {3195} (\bibinfo {year} {1996})}\BibitemShut
  {NoStop}%
\bibitem [{\citenamefont {Carrington}\ \emph {et~al.}(1999)\citenamefont
  {Carrington}, \citenamefont {Giannetta}, \citenamefont {Kim},\ and\
  \citenamefont {Giapintzakis}}]{Car99}%
  \BibitemOpen
  \bibfield  {author} {\bibinfo {author} {\bibfnamefont {A.}~\bibnamefont
  {Carrington}}, \bibinfo {author} {\bibfnamefont {R.~W.}\ \bibnamefont
  {Giannetta}}, \bibinfo {author} {\bibfnamefont {J.~T.}\ \bibnamefont {Kim}},
  \ and\ \bibinfo {author} {\bibfnamefont {J.}~\bibnamefont {Giapintzakis}},\
  }\href@noop {} {\bibfield  {journal} {\bibinfo  {journal} {Phys. Rev. B}\
  }\textbf {\bibinfo {volume} {59}},\ \bibinfo {pages} {R14173} (\bibinfo
  {year} {1999})}\BibitemShut {NoStop}%
\bibitem [{\citenamefont {White}\ \emph {et~al.}(2011)\citenamefont {White},
  \citenamefont {Heslop}, \citenamefont {Holmes}, \citenamefont {Forgan},
  \citenamefont {Hinkov}, \citenamefont {Egetenmeyer}, \citenamefont
  {Gavilano}, \citenamefont {Laver}, \citenamefont {Dewhurst}, \citenamefont
  {Cubitt},\ and\ \citenamefont {Erb}}]{Whi11}%
  \BibitemOpen
  \bibfield  {author} {\bibinfo {author} {\bibfnamefont {J.~S.}\ \bibnamefont
  {White}}, \bibinfo {author} {\bibfnamefont {R.~W.}\ \bibnamefont {Heslop}},
  \bibinfo {author} {\bibfnamefont {A.~T.}\ \bibnamefont {Holmes}}, \bibinfo
  {author} {\bibfnamefont {E.~M.}\ \bibnamefont {Forgan}}, \bibinfo {author}
  {\bibfnamefont {V.}~\bibnamefont {Hinkov}}, \bibinfo {author} {\bibfnamefont
  {N.}~\bibnamefont {Egetenmeyer}}, \bibinfo {author} {\bibfnamefont {J.~L.}\
  \bibnamefont {Gavilano}}, \bibinfo {author} {\bibfnamefont {M.}~\bibnamefont
  {Laver}}, \bibinfo {author} {\bibfnamefont {C.~D.}\ \bibnamefont {Dewhurst}},
  \bibinfo {author} {\bibfnamefont {R.}~\bibnamefont {Cubitt}}, \ and\ \bibinfo
  {author} {\bibfnamefont {A.}~\bibnamefont {Erb}},\ }\href@noop {} {\bibfield
  {journal} {\bibinfo  {journal} {Phys. Rev. B}\ }\textbf {\bibinfo {volume}
  {84}},\ \bibinfo {pages} {104519} (\bibinfo {year} {2011})}\BibitemShut
  {NoStop}%
\bibitem [{\citenamefont {Xiang}\ and\ \citenamefont {Wheatley}(1996)}]{Xia96}%
  \BibitemOpen
  \bibfield  {author} {\bibinfo {author} {\bibfnamefont {T.}~\bibnamefont
  {Xiang}}\ and\ \bibinfo {author} {\bibfnamefont {J.~M.}\ \bibnamefont
  {Wheatley}},\ }\href@noop {} {\bibfield  {journal} {\bibinfo  {journal}
  {Phys. Rev. Lett.}\ }\textbf {\bibinfo {volume} {76}},\ \bibinfo {pages}
  {134} (\bibinfo {year} {1996})}\BibitemShut {NoStop}%
\bibitem [{\citenamefont {Khasanov}\ \emph {et~al.}(2008)\citenamefont
  {Khasanov}, \citenamefont {Shengelaya}, \citenamefont {Karpinski},
  \citenamefont {Bussmann-Holder}, \citenamefont {Keller},\ and\ \citenamefont
  {Müller}}]{Kha08}%
  \BibitemOpen
  \bibfield  {author} {\bibinfo {author} {\bibfnamefont {R.}~\bibnamefont
  {Khasanov}}, \bibinfo {author} {\bibfnamefont {A.}~\bibnamefont
  {Shengelaya}}, \bibinfo {author} {\bibfnamefont {J.}~\bibnamefont
  {Karpinski}}, \bibinfo {author} {\bibfnamefont {A.}~\bibnamefont
  {Bussmann-Holder}}, \bibinfo {author} {\bibfnamefont {H.}~\bibnamefont
  {Keller}}, \ and\ \bibinfo {author} {\bibfnamefont {K.}~\bibnamefont
  {Müller}},\ }\href@noop {} {\bibfield  {journal} {\bibinfo  {journal} {J.
  Supercond. Nov. Magn.}\ }\textbf {\bibinfo {volume} {21}},\ \bibinfo {pages}
  {81} (\bibinfo {year} {2008})}\BibitemShut {NoStop}%
\bibitem [{\citenamefont {Serafin}\ \emph {et~al.}(2010)\citenamefont
  {Serafin}, \citenamefont {Fletcher}, \citenamefont {Adachi}, \citenamefont
  {Hussey},\ and\ \citenamefont {Carrington}}]{Ser10}%
  \BibitemOpen
  \bibfield  {author} {\bibinfo {author} {\bibfnamefont {A.}~\bibnamefont
  {Serafin}}, \bibinfo {author} {\bibfnamefont {J.~D.}\ \bibnamefont
  {Fletcher}}, \bibinfo {author} {\bibfnamefont {S.}~\bibnamefont {Adachi}},
  \bibinfo {author} {\bibfnamefont {N.~E.}\ \bibnamefont {Hussey}}, \ and\
  \bibinfo {author} {\bibfnamefont {A.}~\bibnamefont {Carrington}},\
  }\href@noop {} {\bibfield  {journal} {\bibinfo  {journal} {Phys. Rev. B}\
  }\textbf {\bibinfo {volume} {82}},\ \bibinfo {pages} {140506(R)} (\bibinfo
  {year} {2010})}\BibitemShut {NoStop}%
\bibitem [{\citenamefont {Johnson}\ \emph {et~al.}(1999)\citenamefont
  {Johnson}, \citenamefont {Forgan}, \citenamefont {Lloyd}, \citenamefont
  {Aegerter}, \citenamefont {Lee}, \citenamefont {Cubitt}, \citenamefont
  {Kealey}, \citenamefont {Ager}, \citenamefont {Tajima}, \citenamefont
  {Rykov},\ and\ \citenamefont {Paul}}]{Joh99}%
  \BibitemOpen
  \bibfield  {author} {\bibinfo {author} {\bibfnamefont {S.~T.}\ \bibnamefont
  {Johnson}}, \bibinfo {author} {\bibfnamefont {E.~M.}\ \bibnamefont {Forgan}},
  \bibinfo {author} {\bibfnamefont {S.~H.}\ \bibnamefont {Lloyd}}, \bibinfo
  {author} {\bibfnamefont {C.~M.}\ \bibnamefont {Aegerter}}, \bibinfo {author}
  {\bibfnamefont {S.~L.}\ \bibnamefont {Lee}}, \bibinfo {author} {\bibfnamefont
  {R.}~\bibnamefont {Cubitt}}, \bibinfo {author} {\bibfnamefont {P.~G.}\
  \bibnamefont {Kealey}}, \bibinfo {author} {\bibfnamefont {C.}~\bibnamefont
  {Ager}}, \bibinfo {author} {\bibfnamefont {S.}~\bibnamefont {Tajima}},
  \bibinfo {author} {\bibfnamefont {A.}~\bibnamefont {Rykov}}, \ and\ \bibinfo
  {author} {\bibfnamefont {D.~M.}\ \bibnamefont {Paul}},\ }\href@noop {}
  {\bibfield  {journal} {\bibinfo  {journal} {Phys. Rev. Lett.}\ }\textbf
  {\bibinfo {volume} {82}},\ \bibinfo {pages} {2792} (\bibinfo {year}
  {1999})}\BibitemShut {NoStop}%
\bibitem [{\citenamefont {Brown}\ \emph {et~al.}(2004)\citenamefont {Brown},
  \citenamefont {Charalambous}, \citenamefont {Jones}, \citenamefont {Forgan},
  \citenamefont {Kealey}, \citenamefont {Erb},\ and\ \citenamefont
  {Kohlbrecher}}]{Bro04}%
  \BibitemOpen
  \bibfield  {author} {\bibinfo {author} {\bibfnamefont {S.~P.}\ \bibnamefont
  {Brown}}, \bibinfo {author} {\bibfnamefont {D.}~\bibnamefont {Charalambous}},
  \bibinfo {author} {\bibfnamefont {E.~C.}\ \bibnamefont {Jones}}, \bibinfo
  {author} {\bibfnamefont {E.~M.}\ \bibnamefont {Forgan}}, \bibinfo {author}
  {\bibfnamefont {P.~G.}\ \bibnamefont {Kealey}}, \bibinfo {author}
  {\bibfnamefont {A.}~\bibnamefont {Erb}}, \ and\ \bibinfo {author}
  {\bibfnamefont {J.}~\bibnamefont {Kohlbrecher}},\ }\href@noop {} {\bibfield
  {journal} {\bibinfo  {journal} {Phys. Rev. Lett.}\ }\textbf {\bibinfo
  {volume} {92}},\ \bibinfo {pages} {067004} (\bibinfo {year}
  {2004})}\BibitemShut {NoStop}%
\bibitem [{\citenamefont {White}\ \emph {et~al.}(2008)\citenamefont {White},
  \citenamefont {Brown}, \citenamefont {Forgan}, \citenamefont {Laver},
  \citenamefont {Bowell}, \citenamefont {Lycett}, \citenamefont {Charalambous},
  \citenamefont {Hinkov}, \citenamefont {Erb},\ and\ \citenamefont
  {Kohlbrecher}}]{Whi08}%
  \BibitemOpen
  \bibfield  {author} {\bibinfo {author} {\bibfnamefont {J.~S.}\ \bibnamefont
  {White}}, \bibinfo {author} {\bibfnamefont {S.~P.}\ \bibnamefont {Brown}},
  \bibinfo {author} {\bibfnamefont {E.~M.}\ \bibnamefont {Forgan}}, \bibinfo
  {author} {\bibfnamefont {M.}~\bibnamefont {Laver}}, \bibinfo {author}
  {\bibfnamefont {C.~J.}\ \bibnamefont {Bowell}}, \bibinfo {author}
  {\bibfnamefont {R.~J.}\ \bibnamefont {Lycett}}, \bibinfo {author}
  {\bibfnamefont {D.}~\bibnamefont {Charalambous}}, \bibinfo {author}
  {\bibfnamefont {V.}~\bibnamefont {Hinkov}}, \bibinfo {author} {\bibfnamefont
  {A.}~\bibnamefont {Erb}}, \ and\ \bibinfo {author} {\bibfnamefont
  {J.}~\bibnamefont {Kohlbrecher}},\ }\href@noop {} {\bibfield  {journal}
  {\bibinfo  {journal} {Phys. Rev. B}\ }\textbf {\bibinfo {volume} {78}},\
  \bibinfo {pages} {174513} (\bibinfo {year} {2008})}\BibitemShut {NoStop}%
\bibitem [{\citenamefont {White}\ \emph {et~al.}(2009)\citenamefont {White},
  \citenamefont {Hinkov}, \citenamefont {Heslop}, \citenamefont {Lycett},
  \citenamefont {Forgan}, \citenamefont {Bowell}, \citenamefont {Str\"assle},
  \citenamefont {Abrahamsen}, \citenamefont {Laver}, \citenamefont {Dewhurst},
  \citenamefont {Kohlbrecher}, \citenamefont {Gavilano}, \citenamefont {Mesot},
  \citenamefont {Keimer},\ and\ \citenamefont {Erb}}]{Whi09}%
  \BibitemOpen
  \bibfield  {author} {\bibinfo {author} {\bibfnamefont {J.~S.}\ \bibnamefont
  {White}}, \bibinfo {author} {\bibfnamefont {V.}~\bibnamefont {Hinkov}},
  \bibinfo {author} {\bibfnamefont {R.~W.}\ \bibnamefont {Heslop}}, \bibinfo
  {author} {\bibfnamefont {R.~J.}\ \bibnamefont {Lycett}}, \bibinfo {author}
  {\bibfnamefont {E.~M.}\ \bibnamefont {Forgan}}, \bibinfo {author}
  {\bibfnamefont {C.}~\bibnamefont {Bowell}}, \bibinfo {author} {\bibfnamefont
  {S.}~\bibnamefont {Str\"assle}}, \bibinfo {author} {\bibfnamefont {A.~B.}\
  \bibnamefont {Abrahamsen}}, \bibinfo {author} {\bibfnamefont
  {M.}~\bibnamefont {Laver}}, \bibinfo {author} {\bibfnamefont {C.~D.}\
  \bibnamefont {Dewhurst}}, \bibinfo {author} {\bibfnamefont {J.}~\bibnamefont
  {Kohlbrecher}}, \bibinfo {author} {\bibfnamefont {J.~L.}\ \bibnamefont
  {Gavilano}}, \bibinfo {author} {\bibfnamefont {J.}~\bibnamefont {Mesot}},
  \bibinfo {author} {\bibfnamefont {B.}~\bibnamefont {Keimer}}, \ and\ \bibinfo
  {author} {\bibfnamefont {A.}~\bibnamefont {Erb}},\ }\href@noop {} {\bibfield
  {journal} {\bibinfo  {journal} {Phys. Rev. Lett.}\ }\textbf {\bibinfo
  {volume} {102}},\ \bibinfo {pages} {097001} (\bibinfo {year}
  {2009})}\BibitemShut {NoStop}%
\bibitem [{\citenamefont {Karpinski}\ \emph {et~al.}(1999)\citenamefont
  {Karpinski}, \citenamefont {Meijer}, \citenamefont {Schwer}, \citenamefont
  {Molinski}, \citenamefont {Kopnin}, \citenamefont {Conder}, \citenamefont
  {Angst}, \citenamefont {Jun}, \citenamefont {Kazakov}, \citenamefont
  {Wisniewski}, \citenamefont {Puzniak}, \citenamefont {Hofer}, \citenamefont
  {Alyoshin},\ and\ \citenamefont {Sin}}]{Kar99}%
  \BibitemOpen
  \bibfield  {author} {\bibinfo {author} {\bibfnamefont {J.}~\bibnamefont
  {Karpinski}}, \bibinfo {author} {\bibfnamefont {G.~I.}\ \bibnamefont
  {Meijer}}, \bibinfo {author} {\bibfnamefont {H.}~\bibnamefont {Schwer}},
  \bibinfo {author} {\bibfnamefont {R.}~\bibnamefont {Molinski}}, \bibinfo
  {author} {\bibfnamefont {E.}~\bibnamefont {Kopnin}}, \bibinfo {author}
  {\bibfnamefont {K.}~\bibnamefont {Conder}}, \bibinfo {author} {\bibfnamefont
  {M.}~\bibnamefont {Angst}}, \bibinfo {author} {\bibfnamefont
  {J.}~\bibnamefont {Jun}}, \bibinfo {author} {\bibfnamefont {S.}~\bibnamefont
  {Kazakov}}, \bibinfo {author} {\bibfnamefont {A.}~\bibnamefont {Wisniewski}},
  \bibinfo {author} {\bibfnamefont {R.}~\bibnamefont {Puzniak}}, \bibinfo
  {author} {\bibfnamefont {J.}~\bibnamefont {Hofer}}, \bibinfo {author}
  {\bibfnamefont {V.}~\bibnamefont {Alyoshin}}, \ and\ \bibinfo {author}
  {\bibfnamefont {A.}~\bibnamefont {Sin}},\ }\href@noop {} {\bibfield
  {journal} {\bibinfo  {journal} {Supercond. Sci. Technol.}\ }\textbf {\bibinfo
  {volume} {12}},\ \bibinfo {pages} {R153} (\bibinfo {year}
  {1999})}\BibitemShut {NoStop}%
\bibitem [{\citenamefont {Kogan}(1981)}]{Kog81}%
  \BibitemOpen
  \bibfield  {author} {\bibinfo {author} {\bibfnamefont {V.~G.}\ \bibnamefont
  {Kogan}},\ }\href@noop {} {\bibfield  {journal} {\bibinfo  {journal} {Phys.
  Rev. B}\ }\textbf {\bibinfo {volume} {24}},\ \bibinfo {pages} {1572}
  (\bibinfo {year} {1981})}\BibitemShut {NoStop}%
\bibitem [{\citenamefont {Campbell}\ \emph {et~al.}(1988)\citenamefont
  {Campbell}, \citenamefont {Doria},\ and\ \citenamefont {Kogan}}]{Cam88}%
  \BibitemOpen
  \bibfield  {author} {\bibinfo {author} {\bibfnamefont {L.~J.}\ \bibnamefont
  {Campbell}}, \bibinfo {author} {\bibfnamefont {M.~M.}\ \bibnamefont {Doria}},
  \ and\ \bibinfo {author} {\bibfnamefont {V.~G.}\ \bibnamefont {Kogan}},\
  }\href@noop {} {\bibfield  {journal} {\bibinfo  {journal} {Phys. Rev. B}\
  }\textbf {\bibinfo {volume} {38}},\ \bibinfo {pages} {2439} (\bibinfo {year}
  {1988})}\BibitemShut {NoStop}%
\bibitem [{\citenamefont {Thiemann}\ \emph {et~al.}(1989)\citenamefont
  {Thiemann}, \citenamefont {Radovi\'{c}},\ and\ \citenamefont
  {Kogan}}]{Thi89}%
  \BibitemOpen
  \bibfield  {author} {\bibinfo {author} {\bibfnamefont {S.~L.}\ \bibnamefont
  {Thiemann}}, \bibinfo {author} {\bibfnamefont {Z.}~\bibnamefont
  {Radovi\'{c}}}, \ and\ \bibinfo {author} {\bibfnamefont {V.~G.}\ \bibnamefont
  {Kogan}},\ }\href@noop {} {\bibfield  {journal} {\bibinfo  {journal} {Phys.
  Rev. B}\ }\textbf {\bibinfo {volume} {39}},\ \bibinfo {pages} {11406}
  (\bibinfo {year} {1989})}\BibitemShut {NoStop}%
\bibitem [{\citenamefont {Kogan}\ \emph {et~al.}(1997)\citenamefont {Kogan},
  \citenamefont {Bullock}, \citenamefont {Harmon}, \citenamefont
  {Miranovi\'{c}}, \citenamefont {Dobrosavljevi\'{c}-Gruji\'{c}}, \citenamefont
  {Gammel},\ and\ \citenamefont {Bishop}}]{Kog97a}%
  \BibitemOpen
  \bibfield  {author} {\bibinfo {author} {\bibfnamefont {V.~G.}\ \bibnamefont
  {Kogan}}, \bibinfo {author} {\bibfnamefont {M.}~\bibnamefont {Bullock}},
  \bibinfo {author} {\bibfnamefont {B.}~\bibnamefont {Harmon}}, \bibinfo
  {author} {\bibfnamefont {P.}~\bibnamefont {Miranovi\'{c}}}, \bibinfo {author}
  {\bibfnamefont {L.}~\bibnamefont {Dobrosavljevi\'{c}-Gruji\'{c}}}, \bibinfo
  {author} {\bibfnamefont {P.~L.}\ \bibnamefont {Gammel}}, \ and\ \bibinfo
  {author} {\bibfnamefont {D.~J.}\ \bibnamefont {Bishop}},\ }\href@noop {}
  {\bibfield  {journal} {\bibinfo  {journal} {Phys. Rev. B}\ }\textbf {\bibinfo
  {volume} {55}},\ \bibinfo {pages} {R8693} (\bibinfo {year}
  {1997})}\BibitemShut {NoStop}%
\bibitem [{\citenamefont {Franz}\ \emph {et~al.}(1997)\citenamefont {Franz},
  \citenamefont {Affleck},\ and\ \citenamefont {Amin}}]{Fra97}%
  \BibitemOpen
  \bibfield  {author} {\bibinfo {author} {\bibfnamefont {M.}~\bibnamefont
  {Franz}}, \bibinfo {author} {\bibfnamefont {I.}~\bibnamefont {Affleck}}, \
  and\ \bibinfo {author} {\bibfnamefont {M.~H.~S.}\ \bibnamefont {Amin}},\
  }\href@noop {} {\bibfield  {journal} {\bibinfo  {journal} {Phys. Rev. Lett.}\
  }\textbf {\bibinfo {volume} {79}},\ \bibinfo {pages} {1555} (\bibinfo {year}
  {1997})}\BibitemShut {NoStop}%
\bibitem [{\citenamefont {Ichioka}\ \emph {et~al.}(1999)\citenamefont
  {Ichioka}, \citenamefont {Hasegawa},\ and\ \citenamefont {Machida}}]{Ich99}%
  \BibitemOpen
  \bibfield  {author} {\bibinfo {author} {\bibfnamefont {M.}~\bibnamefont
  {Ichioka}}, \bibinfo {author} {\bibfnamefont {A.}~\bibnamefont {Hasegawa}}, \
  and\ \bibinfo {author} {\bibfnamefont {K.}~\bibnamefont {Machida}},\
  }\href@noop {} {\bibfield  {journal} {\bibinfo  {journal} {Phys. Rev. B}\
  }\textbf {\bibinfo {volume} {59}},\ \bibinfo {pages} {8902} (\bibinfo {year}
  {1999})}\BibitemShut {NoStop}%
\bibitem [{\citenamefont {Nakai}\ \emph {et~al.}(2002)\citenamefont {Nakai},
  \citenamefont {Miranovi\'{c}}, \citenamefont {Ichioka},\ and\ \citenamefont
  {Machida}}]{Nak02}%
  \BibitemOpen
  \bibfield  {author} {\bibinfo {author} {\bibfnamefont {N.}~\bibnamefont
  {Nakai}}, \bibinfo {author} {\bibfnamefont {P.}~\bibnamefont
  {Miranovi\'{c}}}, \bibinfo {author} {\bibfnamefont {M.}~\bibnamefont
  {Ichioka}}, \ and\ \bibinfo {author} {\bibfnamefont {K.}~\bibnamefont
  {Machida}},\ }\href@noop {} {\bibfield  {journal} {\bibinfo  {journal} {Phys.
  Rev. Lett.}\ }\textbf {\bibinfo {volume} {89}},\ \bibinfo {pages} {237004}
  (\bibinfo {year} {2002})}\BibitemShut {NoStop}%
\bibitem [{\citenamefont {Christen}\ \emph {et~al.}(1977)\citenamefont
  {Christen}, \citenamefont {Tasset}, \citenamefont {Spooner},\ and\
  \citenamefont {Mook}}]{Chr77}%
  \BibitemOpen
  \bibfield  {author} {\bibinfo {author} {\bibfnamefont {D.~K.}\ \bibnamefont
  {Christen}}, \bibinfo {author} {\bibfnamefont {F.}~\bibnamefont {Tasset}},
  \bibinfo {author} {\bibfnamefont {S.}~\bibnamefont {Spooner}}, \ and\
  \bibinfo {author} {\bibfnamefont {H.~A.}\ \bibnamefont {Mook}},\ }\href@noop
  {} {\bibfield  {journal} {\bibinfo  {journal} {Phys. Rev. B}\ }\textbf
  {\bibinfo {volume} {15}},\ \bibinfo {pages} {4506} (\bibinfo {year}
  {1977})}\BibitemShut {NoStop}%
\bibitem [{\citenamefont {Yaouanc}\ \emph {et~al.}(1997)\citenamefont
  {Yaouanc}, \citenamefont {Dalmas~de R\'{e}otier},\ and\ \citenamefont
  {Brandt}}]{Yao97}%
  \BibitemOpen
  \bibfield  {author} {\bibinfo {author} {\bibfnamefont {A.}~\bibnamefont
  {Yaouanc}}, \bibinfo {author} {\bibfnamefont {P.}~\bibnamefont {Dalmas~de
  R\'{e}otier}}, \ and\ \bibinfo {author} {\bibfnamefont {E.~H.}\ \bibnamefont
  {Brandt}},\ }\href@noop {} {\bibfield  {journal} {\bibinfo  {journal} {Phys.
  Rev. B}\ }\textbf {\bibinfo {volume} {55}},\ \bibinfo {pages} {11107}
  (\bibinfo {year} {1997})}\BibitemShut {NoStop}%
\bibitem [{\citenamefont {Eskildsen}(2011)}]{Esk11}%
  \BibitemOpen
  \bibfield  {author} {\bibinfo {author} {\bibfnamefont {M.~R.}\ \bibnamefont
  {Eskildsen}},\ }\href@noop {} {\bibfield  {journal} {\bibinfo  {journal}
  {Front. Phys.}\ }\textbf {\bibinfo {volume} {6}},\ \bibinfo {pages} {398}
  (\bibinfo {year} {2011})}\BibitemShut {NoStop}%
\bibitem [{\citenamefont {Chang}\ \emph
  {et~al.}(2012{\natexlab{a}})\citenamefont {Chang}, \citenamefont
  {Doiron-Leyraud}, \citenamefont {Cyr-Choinière}, \citenamefont
  {Grissonnanche}, \citenamefont {Laliberté}, \citenamefont {Hassinger},
  \citenamefont {Reid}, \citenamefont {Daou}, \citenamefont {Pyon},
  \citenamefont {Takayama}, \citenamefont {Takagi},\ and\ \citenamefont
  {Taillefer}}]{Cha12c}%
  \BibitemOpen
  \bibfield  {author} {\bibinfo {author} {\bibfnamefont {J.}~\bibnamefont
  {Chang}}, \bibinfo {author} {\bibfnamefont {N.}~\bibnamefont
  {Doiron-Leyraud}}, \bibinfo {author} {\bibfnamefont {O.}~\bibnamefont
  {Cyr-Choinière}}, \bibinfo {author} {\bibfnamefont {G.}~\bibnamefont
  {Grissonnanche}}, \bibinfo {author} {\bibfnamefont {F.}~\bibnamefont
  {Laliberté}}, \bibinfo {author} {\bibfnamefont {E.}~\bibnamefont
  {Hassinger}}, \bibinfo {author} {\bibfnamefont {J.-P.}\ \bibnamefont {Reid}},
  \bibinfo {author} {\bibfnamefont {R.}~\bibnamefont {Daou}}, \bibinfo {author}
  {\bibfnamefont {S.}~\bibnamefont {Pyon}}, \bibinfo {author} {\bibfnamefont
  {T.}~\bibnamefont {Takayama}}, \bibinfo {author} {\bibfnamefont
  {H.}~\bibnamefont {Takagi}}, \ and\ \bibinfo {author} {\bibfnamefont
  {L.}~\bibnamefont {Taillefer}},\ }\href@noop {} {\bibfield  {journal}
  {\bibinfo  {journal} {Nat. Phys.}\ }\textbf {\bibinfo {volume} {8}},\
  \bibinfo {pages} {751} (\bibinfo {year} {2012}{\natexlab{a}})}\BibitemShut
  {NoStop}%
\bibitem [{\citenamefont {Ramshaw}\ \emph {et~al.}(2012)\citenamefont
  {Ramshaw}, \citenamefont {Day}, \citenamefont {Vignolle}, \citenamefont
  {LeBoeuf}, \citenamefont {Dosanjh}, \citenamefont {Proust}, \citenamefont
  {Taillefer}, \citenamefont {Liang}, \citenamefont {Hardy},\ and\
  \citenamefont {Bonn}}]{Ram12}%
  \BibitemOpen
  \bibfield  {author} {\bibinfo {author} {\bibfnamefont {B.~J.}\ \bibnamefont
  {Ramshaw}}, \bibinfo {author} {\bibfnamefont {J.}~\bibnamefont {Day}},
  \bibinfo {author} {\bibfnamefont {B.}~\bibnamefont {Vignolle}}, \bibinfo
  {author} {\bibfnamefont {D.}~\bibnamefont {LeBoeuf}}, \bibinfo {author}
  {\bibfnamefont {P.}~\bibnamefont {Dosanjh}}, \bibinfo {author} {\bibfnamefont
  {C.}~\bibnamefont {Proust}}, \bibinfo {author} {\bibfnamefont
  {L.}~\bibnamefont {Taillefer}}, \bibinfo {author} {\bibfnamefont
  {R.}~\bibnamefont {Liang}}, \bibinfo {author} {\bibfnamefont {W.~N.}\
  \bibnamefont {Hardy}}, \ and\ \bibinfo {author} {\bibfnamefont {D.~A.}\
  \bibnamefont {Bonn}},\ }\href@noop {} {\bibfield  {journal} {\bibinfo
  {journal} {Phys. Rev. B}\ }\textbf {\bibinfo {volume} {86}},\ \bibinfo
  {pages} {174501} (\bibinfo {year} {2012})}\BibitemShut {NoStop}%
\bibitem [{Not({\natexlab{a}})}]{Not3}%
  \BibitemOpen
  \href@noop {} {} ({\natexlab{a}}),\ \bibinfo {note} {the simplified band
  structure neglects an explicit SC chain band, but we expect $\lambda_{a}(T)$
  to be insensitive to any superfluid density associated with the CuO chain
  states. This assumption is not necessarily true for the $T$-variation of
  $\xi_{b}(T)$. However, at such a low applied field the exponential term in
  Eq.~2 that is dependent on $\xi_{b}(T)$ remains close to 1 (within
  $\sim$4~\%), and so the $T$-dependent behavior of $|F\left(\textrm{\textbf{q}
  horiz}\right)|$ is dominated by $\lambda_{a}(T)$.}\BibitemShut {Stop}%
\bibitem [{\citenamefont {Sok}\ \emph {et~al.}(1995)\citenamefont {Sok},
  \citenamefont {Xu}, \citenamefont {Chen}, \citenamefont {Suh}, \citenamefont
  {Gohng}, \citenamefont {Finnemore}, \citenamefont {Kramer}, \citenamefont
  {Schwartzkopf},\ and\ \citenamefont {Dabrowski}}]{Sok95}%
  \BibitemOpen
  \bibfield  {author} {\bibinfo {author} {\bibfnamefont {J.}~\bibnamefont
  {Sok}}, \bibinfo {author} {\bibfnamefont {M.}~\bibnamefont {Xu}}, \bibinfo
  {author} {\bibfnamefont {W.}~\bibnamefont {Chen}}, \bibinfo {author}
  {\bibfnamefont {B.~J.}\ \bibnamefont {Suh}}, \bibinfo {author} {\bibfnamefont
  {J.}~\bibnamefont {Gohng}}, \bibinfo {author} {\bibfnamefont {D.~K.}\
  \bibnamefont {Finnemore}}, \bibinfo {author} {\bibfnamefont {M.~J.}\
  \bibnamefont {Kramer}}, \bibinfo {author} {\bibfnamefont {L.~A.}\
  \bibnamefont {Schwartzkopf}}, \ and\ \bibinfo {author} {\bibfnamefont
  {B.}~\bibnamefont {Dabrowski}},\ }\href@noop {} {\bibfield  {journal}
  {\bibinfo  {journal} {Phys. Rev. B}\ }\textbf {\bibinfo {volume} {51}},\
  \bibinfo {pages} {6035} (\bibinfo {year} {1995})}\BibitemShut {NoStop}%
\bibitem [{\citenamefont {Kondo}\ \emph {et~al.}(2010)\citenamefont {Kondo},
  \citenamefont {Khasanov}, \citenamefont {Karpinski}, \citenamefont {Kazakov},
  \citenamefont {Zhigadlo}, \citenamefont {Bukowski}, \citenamefont {Shi},
  \citenamefont {Bendounan}, \citenamefont {Sassa}, \citenamefont {Chang},
  \citenamefont {Pailh\'es}, \citenamefont {Mesot}, \citenamefont {Schmalian},
  \citenamefont {Keller},\ and\ \citenamefont {Kaminski}}]{Kon10}%
  \BibitemOpen
  \bibfield  {author} {\bibinfo {author} {\bibfnamefont {T.}~\bibnamefont
  {Kondo}}, \bibinfo {author} {\bibfnamefont {R.}~\bibnamefont {Khasanov}},
  \bibinfo {author} {\bibfnamefont {J.}~\bibnamefont {Karpinski}}, \bibinfo
  {author} {\bibfnamefont {S.~M.}\ \bibnamefont {Kazakov}}, \bibinfo {author}
  {\bibfnamefont {N.~D.}\ \bibnamefont {Zhigadlo}}, \bibinfo {author}
  {\bibfnamefont {Z.}~\bibnamefont {Bukowski}}, \bibinfo {author}
  {\bibfnamefont {M.}~\bibnamefont {Shi}}, \bibinfo {author} {\bibfnamefont
  {A.}~\bibnamefont {Bendounan}}, \bibinfo {author} {\bibfnamefont
  {Y.}~\bibnamefont {Sassa}}, \bibinfo {author} {\bibfnamefont
  {J.}~\bibnamefont {Chang}}, \bibinfo {author} {\bibfnamefont
  {S.}~\bibnamefont {Pailh\'es}}, \bibinfo {author} {\bibfnamefont
  {J.}~\bibnamefont {Mesot}}, \bibinfo {author} {\bibfnamefont
  {J.}~\bibnamefont {Schmalian}}, \bibinfo {author} {\bibfnamefont
  {H.}~\bibnamefont {Keller}}, \ and\ \bibinfo {author} {\bibfnamefont
  {A.}~\bibnamefont {Kaminski}},\ }\href@noop {} {\bibfield  {journal}
  {\bibinfo  {journal} {Phys. Rev. Lett.}\ }\textbf {\bibinfo {volume} {105}},\
  \bibinfo {pages} {267003} (\bibinfo {year} {2010})}\BibitemShut {NoStop}%
\bibitem [{Not({\natexlab{b}})}]{Not1}%
  \BibitemOpen
  \href@noop {} {} ({\natexlab{b}}),\ \bibinfo {note} {we note also that the
  effects of chain disorder, which can suppress a positive curvature in
  $n_{i}(T)$~\citep{Atk99}, are negated by studies on both Y124 and Y1237 which
  each have filled chains.}\BibitemShut {Stop}%
\bibitem [{\citenamefont {Doiron-Leyraud}\ \emph {et~al.}(2007)\citenamefont
  {Doiron-Leyraud}, \citenamefont {Proust}, \citenamefont {LeBoeuf},
  \citenamefont {Levallois}, \citenamefont {Bonnemaison}, \citenamefont
  {Liang}, \citenamefont {Bonn}, \citenamefont {Hardy},\ and\ \citenamefont
  {Taillefer}}]{Doi07}%
  \BibitemOpen
  \bibfield  {author} {\bibinfo {author} {\bibfnamefont {N.}~\bibnamefont
  {Doiron-Leyraud}}, \bibinfo {author} {\bibfnamefont {C.}~\bibnamefont
  {Proust}}, \bibinfo {author} {\bibfnamefont {D.}~\bibnamefont {LeBoeuf}},
  \bibinfo {author} {\bibfnamefont {J.}~\bibnamefont {Levallois}}, \bibinfo
  {author} {\bibfnamefont {J.-B.}\ \bibnamefont {Bonnemaison}}, \bibinfo
  {author} {\bibfnamefont {R.}~\bibnamefont {Liang}}, \bibinfo {author}
  {\bibfnamefont {D.~A.}\ \bibnamefont {Bonn}}, \bibinfo {author}
  {\bibfnamefont {W.~N.}\ \bibnamefont {Hardy}}, \ and\ \bibinfo {author}
  {\bibfnamefont {L.}~\bibnamefont {Taillefer}},\ }\href@noop {} {\bibfield
  {journal} {\bibinfo  {journal} {Nature (London)}\ }\textbf {\bibinfo {volume}
  {447}},\ \bibinfo {pages} {565} (\bibinfo {year} {2007})}\BibitemShut
  {NoStop}%
\bibitem [{\citenamefont {Yelland}\ \emph {et~al.}(2008)\citenamefont
  {Yelland}, \citenamefont {Singleton}, \citenamefont {Mielke}, \citenamefont
  {Harrison}, \citenamefont {Balakirev}, \citenamefont {Dabrowski},\ and\
  \citenamefont {Cooper}}]{Yel08}%
  \BibitemOpen
  \bibfield  {author} {\bibinfo {author} {\bibfnamefont {E.~A.}\ \bibnamefont
  {Yelland}}, \bibinfo {author} {\bibfnamefont {J.}~\bibnamefont {Singleton}},
  \bibinfo {author} {\bibfnamefont {C.~H.}\ \bibnamefont {Mielke}}, \bibinfo
  {author} {\bibfnamefont {N.}~\bibnamefont {Harrison}}, \bibinfo {author}
  {\bibfnamefont {F.~F.}\ \bibnamefont {Balakirev}}, \bibinfo {author}
  {\bibfnamefont {B.}~\bibnamefont {Dabrowski}}, \ and\ \bibinfo {author}
  {\bibfnamefont {J.~R.}\ \bibnamefont {Cooper}},\ }\href@noop {} {\bibfield
  {journal} {\bibinfo  {journal} {Phys. Rev. Lett.}\ }\textbf {\bibinfo
  {volume} {100}},\ \bibinfo {pages} {047003} (\bibinfo {year}
  {2008})}\BibitemShut {NoStop}%
\bibitem [{\citenamefont {Bangura}\ \emph {et~al.}(2008)\citenamefont
  {Bangura}, \citenamefont {Fletcher}, \citenamefont {Carrington},
  \citenamefont {Levallois}, \citenamefont {Nardone}, \citenamefont {Vignolle},
  \citenamefont {Heard}, \citenamefont {Doiron-Leyraud}, \citenamefont
  {LeBoeuf}, \citenamefont {Taillefer}, \citenamefont {Adachi}, \citenamefont
  {Proust},\ and\ \citenamefont {Hussey}}]{Ban08}%
  \BibitemOpen
  \bibfield  {author} {\bibinfo {author} {\bibfnamefont {A.~F.}\ \bibnamefont
  {Bangura}}, \bibinfo {author} {\bibfnamefont {J.~D.}\ \bibnamefont
  {Fletcher}}, \bibinfo {author} {\bibfnamefont {A.}~\bibnamefont
  {Carrington}}, \bibinfo {author} {\bibfnamefont {J.}~\bibnamefont
  {Levallois}}, \bibinfo {author} {\bibfnamefont {M.}~\bibnamefont {Nardone}},
  \bibinfo {author} {\bibfnamefont {B.}~\bibnamefont {Vignolle}}, \bibinfo
  {author} {\bibfnamefont {P.~J.}\ \bibnamefont {Heard}}, \bibinfo {author}
  {\bibfnamefont {N.}~\bibnamefont {Doiron-Leyraud}}, \bibinfo {author}
  {\bibfnamefont {D.}~\bibnamefont {LeBoeuf}}, \bibinfo {author} {\bibfnamefont
  {L.}~\bibnamefont {Taillefer}}, \bibinfo {author} {\bibfnamefont
  {S.}~\bibnamefont {Adachi}}, \bibinfo {author} {\bibfnamefont
  {C.}~\bibnamefont {Proust}}, \ and\ \bibinfo {author} {\bibfnamefont {N.~E.}\
  \bibnamefont {Hussey}},\ }\href@noop {} {\bibfield  {journal} {\bibinfo
  {journal} {Phys. Rev. Lett.}\ }\textbf {\bibinfo {volume} {100}},\ \bibinfo
  {pages} {047004} (\bibinfo {year} {2008})}\BibitemShut {NoStop}%
\bibitem [{\citenamefont {Ghiringhelli}\ \emph {et~al.}(2012)\citenamefont
  {Ghiringhelli}, \citenamefont {Le~Tacon}, \citenamefont {Minola},
  \citenamefont {Blanco-Canosa}, \citenamefont {Mazzoli}, \citenamefont
  {Brookes}, \citenamefont {De~Luca}, \citenamefont {Frano}, \citenamefont
  {Hawthorn}, \citenamefont {He}, \citenamefont {Loew}, \citenamefont {Sala},
  \citenamefont {Peets}, \citenamefont {Salluzzo}, \citenamefont {Schierle},
  \citenamefont {Sutarto}, \citenamefont {Sawatzky}, \citenamefont {Weschke},
  \citenamefont {Keimer},\ and\ \citenamefont {Braicovich}}]{Ghi12}%
  \BibitemOpen
  \bibfield  {author} {\bibinfo {author} {\bibfnamefont {G.}~\bibnamefont
  {Ghiringhelli}}, \bibinfo {author} {\bibfnamefont {M.}~\bibnamefont
  {Le~Tacon}}, \bibinfo {author} {\bibfnamefont {M.}~\bibnamefont {Minola}},
  \bibinfo {author} {\bibfnamefont {S.}~\bibnamefont {Blanco-Canosa}}, \bibinfo
  {author} {\bibfnamefont {C.}~\bibnamefont {Mazzoli}}, \bibinfo {author}
  {\bibfnamefont {N.~B.}\ \bibnamefont {Brookes}}, \bibinfo {author}
  {\bibfnamefont {G.~M.}\ \bibnamefont {De~Luca}}, \bibinfo {author}
  {\bibfnamefont {A.}~\bibnamefont {Frano}}, \bibinfo {author} {\bibfnamefont
  {D.~G.}\ \bibnamefont {Hawthorn}}, \bibinfo {author} {\bibfnamefont
  {F.}~\bibnamefont {He}}, \bibinfo {author} {\bibfnamefont {T.}~\bibnamefont
  {Loew}}, \bibinfo {author} {\bibfnamefont {M.~M.}\ \bibnamefont {Sala}},
  \bibinfo {author} {\bibfnamefont {D.~C.}\ \bibnamefont {Peets}}, \bibinfo
  {author} {\bibfnamefont {M.}~\bibnamefont {Salluzzo}}, \bibinfo {author}
  {\bibfnamefont {E.}~\bibnamefont {Schierle}}, \bibinfo {author}
  {\bibfnamefont {R.}~\bibnamefont {Sutarto}}, \bibinfo {author} {\bibfnamefont
  {G.~A.}\ \bibnamefont {Sawatzky}}, \bibinfo {author} {\bibfnamefont
  {E.}~\bibnamefont {Weschke}}, \bibinfo {author} {\bibfnamefont
  {B.}~\bibnamefont {Keimer}}, \ and\ \bibinfo {author} {\bibfnamefont
  {L.}~\bibnamefont {Braicovich}},\ }\href@noop {} {\bibfield  {journal}
  {\bibinfo  {journal} {Science}\ }\textbf {\bibinfo {volume} {337}},\ \bibinfo
  {pages} {821} (\bibinfo {year} {2012})}\BibitemShut {NoStop}%
\bibitem [{\citenamefont {Chang}\ \emph
  {et~al.}(2012{\natexlab{b}})\citenamefont {Chang}, \citenamefont {Blackburn},
  \citenamefont {Holmes}, \citenamefont {Christensen}, \citenamefont {Larsen},
  \citenamefont {Mesot}, \citenamefont {Liang}, \citenamefont {Bonn},
  \citenamefont {Hardy}, \citenamefont {Watenphul}, \citenamefont
  {v.~Zimmermann}, \citenamefont {Forgan},\ and\ \citenamefont
  {Hayden}}]{Cha12b}%
  \BibitemOpen
  \bibfield  {author} {\bibinfo {author} {\bibfnamefont {J.}~\bibnamefont
  {Chang}}, \bibinfo {author} {\bibfnamefont {E.}~\bibnamefont {Blackburn}},
  \bibinfo {author} {\bibfnamefont {A.~T.}\ \bibnamefont {Holmes}}, \bibinfo
  {author} {\bibfnamefont {N.~B.}\ \bibnamefont {Christensen}}, \bibinfo
  {author} {\bibfnamefont {J.}~\bibnamefont {Larsen}}, \bibinfo {author}
  {\bibfnamefont {J.}~\bibnamefont {Mesot}}, \bibinfo {author} {\bibfnamefont
  {R.}~\bibnamefont {Liang}}, \bibinfo {author} {\bibfnamefont {D.~A.}\
  \bibnamefont {Bonn}}, \bibinfo {author} {\bibfnamefont {W.~N.}\ \bibnamefont
  {Hardy}}, \bibinfo {author} {\bibfnamefont {A.}~\bibnamefont {Watenphul}},
  \bibinfo {author} {\bibfnamefont {M.}~\bibnamefont {v.~Zimmermann}}, \bibinfo
  {author} {\bibfnamefont {E.~M.}\ \bibnamefont {Forgan}}, \ and\ \bibinfo
  {author} {\bibfnamefont {S.~M.}\ \bibnamefont {Hayden}},\ }\href@noop {}
  {\bibfield  {journal} {\bibinfo  {journal} {Nat. Phys.}\ }\textbf {\bibinfo
  {volume} {8}},\ \bibinfo {pages} {871} (\bibinfo {year}
  {2012}{\natexlab{b}})}\BibitemShut {NoStop}%
\bibitem [{\citenamefont {Achkar}\ \emph {et~al.}(2012)\citenamefont {Achkar},
  \citenamefont {Sutarto}, \citenamefont {Mao}, \citenamefont {He},
  \citenamefont {Frano}, \citenamefont {Blanco-Canosa}, \citenamefont
  {Le~Tacon}, \citenamefont {Ghiringhelli}, \citenamefont {Braicovich},
  \citenamefont {Minola}, \citenamefont {Moretti~Sala}, \citenamefont
  {Mazzoli}, \citenamefont {Liang}, \citenamefont {Bonn}, \citenamefont
  {Hardy}, \citenamefont {Keimer}, \citenamefont {Sawatzky},\ and\
  \citenamefont {Hawthorn}}]{Ach12}%
  \BibitemOpen
  \bibfield  {author} {\bibinfo {author} {\bibfnamefont {A.~J.}\ \bibnamefont
  {Achkar}}, \bibinfo {author} {\bibfnamefont {R.}~\bibnamefont {Sutarto}},
  \bibinfo {author} {\bibfnamefont {X.}~\bibnamefont {Mao}}, \bibinfo {author}
  {\bibfnamefont {F.}~\bibnamefont {He}}, \bibinfo {author} {\bibfnamefont
  {A.}~\bibnamefont {Frano}}, \bibinfo {author} {\bibfnamefont
  {S.}~\bibnamefont {Blanco-Canosa}}, \bibinfo {author} {\bibfnamefont
  {M.}~\bibnamefont {Le~Tacon}}, \bibinfo {author} {\bibfnamefont
  {G.}~\bibnamefont {Ghiringhelli}}, \bibinfo {author} {\bibfnamefont
  {L.}~\bibnamefont {Braicovich}}, \bibinfo {author} {\bibfnamefont
  {M.}~\bibnamefont {Minola}}, \bibinfo {author} {\bibfnamefont
  {M.}~\bibnamefont {Moretti~Sala}}, \bibinfo {author} {\bibfnamefont
  {C.}~\bibnamefont {Mazzoli}}, \bibinfo {author} {\bibfnamefont
  {R.}~\bibnamefont {Liang}}, \bibinfo {author} {\bibfnamefont {D.~A.}\
  \bibnamefont {Bonn}}, \bibinfo {author} {\bibfnamefont {W.~N.}\ \bibnamefont
  {Hardy}}, \bibinfo {author} {\bibfnamefont {B.}~\bibnamefont {Keimer}},
  \bibinfo {author} {\bibfnamefont {G.~A.}\ \bibnamefont {Sawatzky}}, \ and\
  \bibinfo {author} {\bibfnamefont {D.~G.}\ \bibnamefont {Hawthorn}},\
  }\href@noop {} {\bibfield  {journal} {\bibinfo  {journal} {Phys. Rev. Lett.}\
  }\textbf {\bibinfo {volume} {109}},\ \bibinfo {pages} {167001} (\bibinfo
  {year} {2012})}\BibitemShut {NoStop}%
\bibitem [{\citenamefont {LeBoeuf}\ \emph {et~al.}(2007)\citenamefont
  {LeBoeuf}, \citenamefont {Doiron-Leyraud}, \citenamefont {Levallois},
  \citenamefont {Daou}, \citenamefont {Bonnemaison}, \citenamefont {Hussey},
  \citenamefont {Balicas}, \citenamefont {Ramshaw}, \citenamefont {Liang},
  \citenamefont {Bonn}, \citenamefont {Hardy}, \citenamefont {Adachi},
  \citenamefont {Proust},\ and\ \citenamefont {Taillefer}}]{LeB07}%
  \BibitemOpen
  \bibfield  {author} {\bibinfo {author} {\bibfnamefont {D.}~\bibnamefont
  {LeBoeuf}}, \bibinfo {author} {\bibfnamefont {N.}~\bibnamefont
  {Doiron-Leyraud}}, \bibinfo {author} {\bibfnamefont {J.}~\bibnamefont
  {Levallois}}, \bibinfo {author} {\bibfnamefont {R.}~\bibnamefont {Daou}},
  \bibinfo {author} {\bibfnamefont {J.-B.}\ \bibnamefont {Bonnemaison}},
  \bibinfo {author} {\bibfnamefont {N.~E.}\ \bibnamefont {Hussey}}, \bibinfo
  {author} {\bibfnamefont {L.}~\bibnamefont {Balicas}}, \bibinfo {author}
  {\bibfnamefont {B.~J.}\ \bibnamefont {Ramshaw}}, \bibinfo {author}
  {\bibfnamefont {R.}~\bibnamefont {Liang}}, \bibinfo {author} {\bibfnamefont
  {D.~A.}\ \bibnamefont {Bonn}}, \bibinfo {author} {\bibfnamefont
  {W.}~\bibnamefont {Hardy}}, \bibinfo {author} {\bibfnamefont
  {S.}~\bibnamefont {Adachi}}, \bibinfo {author} {\bibfnamefont
  {C.}~\bibnamefont {Proust}}, \ and\ \bibinfo {author} {\bibfnamefont
  {L.}~\bibnamefont {Taillefer}},\ }\href@noop {} {\bibfield  {journal}
  {\bibinfo  {journal} {Nature}\ }\textbf {\bibinfo {volume} {450}},\ \bibinfo
  {pages} {533} (\bibinfo {year} {2007})}\BibitemShut {NoStop}%
\bibitem [{\citenamefont {Rourke}\ \emph {et~al.}(2010)\citenamefont {Rourke},
  \citenamefont {Bangura}, \citenamefont {Proust}, \citenamefont {Levallois},
  \citenamefont {Doiron-Leyraud}, \citenamefont {LeBoeuf}, \citenamefont
  {Taillefer}, \citenamefont {Adachi}, \citenamefont {Sutherland},\ and\
  \citenamefont {Hussey}}]{Rou10}%
  \BibitemOpen
  \bibfield  {author} {\bibinfo {author} {\bibfnamefont {P.~M.~C.}\
  \bibnamefont {Rourke}}, \bibinfo {author} {\bibfnamefont {A.~F.}\
  \bibnamefont {Bangura}}, \bibinfo {author} {\bibfnamefont {C.}~\bibnamefont
  {Proust}}, \bibinfo {author} {\bibfnamefont {J.}~\bibnamefont {Levallois}},
  \bibinfo {author} {\bibfnamefont {N.}~\bibnamefont {Doiron-Leyraud}},
  \bibinfo {author} {\bibfnamefont {D.}~\bibnamefont {LeBoeuf}}, \bibinfo
  {author} {\bibfnamefont {L.}~\bibnamefont {Taillefer}}, \bibinfo {author}
  {\bibfnamefont {S.}~\bibnamefont {Adachi}}, \bibinfo {author} {\bibfnamefont
  {M.~L.}\ \bibnamefont {Sutherland}}, \ and\ \bibinfo {author} {\bibfnamefont
  {N.~E.}\ \bibnamefont {Hussey}},\ }\href@noop {} {\bibfield  {journal}
  {\bibinfo  {journal} {Phys. Rev. B}\ }\textbf {\bibinfo {volume} {82}},\
  \bibinfo {pages} {020514} (\bibinfo {year} {2010})}\BibitemShut {NoStop}%
\bibitem [{\citenamefont {Chang}\ \emph
  {et~al.}(2012{\natexlab{c}})\citenamefont {Chang}, \citenamefont {White},
  \citenamefont {Laver}, \citenamefont {Bowell}, \citenamefont {Brown},
  \citenamefont {Holmes}, \citenamefont {Maechler}, \citenamefont {Str\"assle},
  \citenamefont {Gilardi}, \citenamefont {Gerber}, \citenamefont {Kurosawa},
  \citenamefont {Momono}, \citenamefont {Oda}, \citenamefont {Ido},
  \citenamefont {Lipscombe}, \citenamefont {Hayden}, \citenamefont {Dewhurst},
  \citenamefont {Vavrin}, \citenamefont {Gavilano}, \citenamefont
  {Kohlbrecher}, \citenamefont {Forgan},\ and\ \citenamefont {Mesot}}]{Cha12}%
  \BibitemOpen
  \bibfield  {author} {\bibinfo {author} {\bibfnamefont {J.}~\bibnamefont
  {Chang}}, \bibinfo {author} {\bibfnamefont {J.~S.}\ \bibnamefont {White}},
  \bibinfo {author} {\bibfnamefont {M.}~\bibnamefont {Laver}}, \bibinfo
  {author} {\bibfnamefont {C.~J.}\ \bibnamefont {Bowell}}, \bibinfo {author}
  {\bibfnamefont {S.~P.}\ \bibnamefont {Brown}}, \bibinfo {author}
  {\bibfnamefont {A.~T.}\ \bibnamefont {Holmes}}, \bibinfo {author}
  {\bibfnamefont {L.}~\bibnamefont {Maechler}}, \bibinfo {author}
  {\bibfnamefont {S.}~\bibnamefont {Str\"assle}}, \bibinfo {author}
  {\bibfnamefont {R.}~\bibnamefont {Gilardi}}, \bibinfo {author} {\bibfnamefont
  {S.}~\bibnamefont {Gerber}}, \bibinfo {author} {\bibfnamefont
  {T.}~\bibnamefont {Kurosawa}}, \bibinfo {author} {\bibfnamefont
  {N.}~\bibnamefont {Momono}}, \bibinfo {author} {\bibfnamefont
  {M.}~\bibnamefont {Oda}}, \bibinfo {author} {\bibfnamefont {M.}~\bibnamefont
  {Ido}}, \bibinfo {author} {\bibfnamefont {O.~J.}\ \bibnamefont {Lipscombe}},
  \bibinfo {author} {\bibfnamefont {S.~M.}\ \bibnamefont {Hayden}}, \bibinfo
  {author} {\bibfnamefont {C.~D.}\ \bibnamefont {Dewhurst}}, \bibinfo {author}
  {\bibfnamefont {R.}~\bibnamefont {Vavrin}}, \bibinfo {author} {\bibfnamefont
  {J.}~\bibnamefont {Gavilano}}, \bibinfo {author} {\bibfnamefont
  {J.}~\bibnamefont {Kohlbrecher}}, \bibinfo {author} {\bibfnamefont {E.~M.}\
  \bibnamefont {Forgan}}, \ and\ \bibinfo {author} {\bibfnamefont
  {J.}~\bibnamefont {Mesot}},\ }\href@noop {} {\bibfield  {journal} {\bibinfo
  {journal} {Phys. Rev. B}\ }\textbf {\bibinfo {volume} {85}},\ \bibinfo
  {pages} {134520} (\bibinfo {year} {2012}{\natexlab{c}})}\BibitemShut
  {NoStop}%
\bibitem [{\citenamefont {Clem}(1975)}]{Cle75}%
  \BibitemOpen
  \bibfield  {author} {\bibinfo {author} {\bibfnamefont {J.~R.}\ \bibnamefont
  {Clem}},\ }\href@noop {} {\bibfield  {journal} {\bibinfo  {journal} {J. Low
  Temp. Phys.}\ }\textbf {\bibinfo {volume} {18}},\ \bibinfo {pages} {427}
  (\bibinfo {year} {1975})}\BibitemShut {NoStop}%
\bibitem [{\citenamefont {Atkinson}(1999)}]{Atk99}%
  \BibitemOpen
  \bibfield  {author} {\bibinfo {author} {\bibfnamefont {W.~A.}\ \bibnamefont
  {Atkinson}},\ }\href@noop {} {\bibfield  {journal} {\bibinfo  {journal}
  {Phys. Rev. B}\ }\textbf {\bibinfo {volume} {59}},\ \bibinfo {pages} {3377}
  (\bibinfo {year} {1999})}\BibitemShut {NoStop}%
\end{thebibliography}%

\end{document}